\documentclass[review,3p,times]{elsarticle}

\usepackage{amsmath,hyperref}
\usepackage{lineno}

\usepackage{epstopdf}
\journal{Springer}

\usepackage{graphicx}
\usepackage{dcolumn}
\usepackage{bm}
\usepackage{epsfig}
\usepackage{booktabs}
\usepackage{subfigure}
\usepackage{graphics}
\usepackage{amssymb}
\usepackage{amsmath}
\usepackage{array}
\usepackage{color}
\usepackage{booktabs}
\usepackage{multirow}
\usepackage{caption}
\usepackage{chngpage}
\usepackage{subfigure}
\usepackage{mathrsfs,gensymb,float}
\usepackage{algorithm, algorithmic}

\biboptions{numbers,sort&compress}

\makeatletter
\newenvironment{breakablealgorithm}
{
	\begin{center}
		\refstepcounter{algorithm}
		\hrule height.8pt depth0pt \kern2pt
		\renewcommand{\caption}[2][\relax]{
			{\raggedright\textbf{\ALG@name~\thealgorithm} ##2\par}%
			\ifx\relax##1\relax 
			\addcontentsline{loa}{algorithm}{\protect\numberline{\thealgorithm}##2}%
			\else 
			\addcontentsline{loa}{algorithm}{\protect\numberline{\thealgorithm}##1}%
			\fi
			\kern2pt\hrule\kern2pt
		}
	}{
		\kern2pt\hrule\relax
	\end{center}
}
\makeatother

\begin{document}

\captionsetup[figure]{labelfont={bf},name={Fig.},labelsep=period}        

\begin{frontmatter}
	
	\title{A thermodynamically consistent phase-field lattice Boltzmann method for two-phase electrohydrodynamic flows}
	\author[1]{Fang Xiong}
	\author[1]{Lei Wang\corref{mycorrespondingauthor}}
	\cortext[mycorrespondingauthor]{Corresponding author}
	\ead{leiwang@cug.edu.cn, wangleir1989@126.com}
	\author[3]{Jiangxu Huang}
	\author[2]{Kang Luo}

\address[1]{School of Mathematics and Physics, China University of Geosciences, Wuhan 430074, China}
\address[3]{School of Mathematics and Statistics, Huazhong University of Science and Technology, Wuhan 430074, China}
\address[2]{School of Energy Science and Engineering, Harbin Institute of Technology, Harbin 150001, China}

\begin{abstract}
In this work, we aim to develop a phase-field based lattice Boltzmann (LB) method for simulating two-phase electrohydrodynamics (EHD) flows, which allows for different properties (densities, viscosities, conductivity and permittivity) of each phase while maintaining thermodynamic consistency. To this end, we first present a theoretical analysis on the two-phase EHD flows by using the Onsager's variational principle, which is an extension of Rayleigh's principle of least energy dissipation and, naturally, guarantees thermodynamic consistency. It shows that the governing equations of the model include the hydrodynamic equations, Cahn-Hilliard equation coupled with additional electrical effect, and the full Poisson-Nernst-Planck electrokinetic equations. After that, a coupled lattice Boltzmann (LB) scheme is constructed for simulating two-phase EHD flows. In particular, in order to handle two-phase EHD flows with a relatively larger electric permittivity ratio, we also introduce a delicately designed discrete forcing term into the LB equation for electrostatic field. Moreover, some numerical examples including two-phase EHD flows in planar layers and charge diffusion of a Gaussian bell are simulated with the developed LB method. It is shown that our numerical scheme shares a second-order convergence rate in space in predicting  electric potential and charge density. Finally, we used the current model to simulate the deformation of a droplet under an electric field and the dynamics of droplet detachment in reversed electrowetting. Our numerical results align well with the theoretic solutions, and the available experimental/numerical data,  demonstrating that the proposed method is feasible for simulating two-phase EHD flows. 

\end{abstract}


\end{frontmatter}

\section{Introduction}
Electrohydrodynamics (EHD) is an interdisciplinary science that deals with the interaction of fluids with electric fields \cite{saville1997ehd}. As one of the fundamental subfields of EHD, a deep understanding of the physics and dynamics of multiphase electrohydrodynamic flow is crucial because of its relevance in a broad range of microfluidics and industrial processes, such as ink-jet printing \cite{raje2014a}, electrospraying \cite{deng2011elec}, manipulation of micro-drops by continuous electrowetting \cite{Bhattacharya2024un},  oil extraction from oil-water emulsions \cite{Vigo2010agg}, EHD pumps \cite{lai2020ehd}, to name a few. In this setting, in order to advance the design and development of the aforementioned technologies and explore new applications, it is crucial to gain a comprehensive understanding of the flow physics related to the multiphase EHD.

Generally, when an electric field is applied to a two-fluid system, a net force is generated at the fluid-fluid interface due to the imbalance of electrical properties between the fluids \cite{Papageorgiou2019film}. For perfectly dielectric or perfectly conducting fluids,  the electric surface force solely acts in the direction normal to the fluid interface, which is further balanced by capillary traction due to surface tension \cite{Landau1975elec,Melcher1975elec}. In such ideal situations, the fluids remain motionless at steady state, and the resultant phenomenon is usually called electrohydrostatics \cite{Melcher1975elec,KONSKI1953the,BASARAN1989axi}. However, as there are no perfectly insulating or perfectly conducting fluids in the real world \cite{Feng1996acom}, researchers have long been interested in investigating fluids with finite electrical conductivity. In the pioneering work of Taylor \cite{Taylor1964Dis}, it was observed that the weakly conducting of the fluids allows electrical charges to accumulate at the droplet surface,  resulting in a tangential electric stress to be generated, which in turn requires a tangential viscous stress to balance it, leading to an EHD flow formed in the system. The EHD theory proposed in Taylor's classical study is known as the leaky dielectric model \cite{saville1997ehd, Melcher1975elec}, which is now widely employed by researchers in theoretical and numerical investigations of droplet deformation under a uniform electric field \cite{Craster2005elec,Mori2018from,Supeene2008defor,Zhang2005A2D,Zhang2021enh,Zhao2023ele}. It shows that Taylor's leaky dielectric model is capable of predicting droplet deformations for small values of capillary number. However, there are some quantitative discrepancies between the leaky dielectric theory and experiment when the droplet undergoes large deformation \cite{Feng1996acom,Feng1999Elec}. To unravel the cause of the discrepancies, in a noteworthy study, Feng et al. \cite{Feng1996acom} investigated the EHD behavior of a drop at finite electric Reynolds number, which defined as the ratio of charge convection to Ohmic conduction. It is found that the nonlinear surface charge convection, which explicitly appears in the leaky-dielectric model but is neglected in the original study, may account for the discrepancies between Taylor's theoretical prediction and experiment. On the basis of Feng et al.'s work \cite{Feng1996acom}, the influence of surface charge convection has garnered significant attention from some researchers in the EHD community \cite{Yariv2016the,Sengupta2017the,Mandal2016effec,Luo2020Num}, and it is now well accepted that incorporating the mechanism of surface charge convection in the corresponding EHD model is more aligned with the actual physical circumstances.  

Apart from the behavior of fluid interfaces under the action of electric fields, understanding droplet dynamics on a solid electrode is also of prime importance due to its wide applications in engineering \cite{Hayes2003video,Eral2011sup}, especially in modern-day microfluidic systems \cite{Paik2003elec,Wang2005elec}. The first work to investigate the effect of an electric field on wetting behavior is attributed to Lippman \cite{Lippmann1875rela}, who found that the presence of a voltage tends to enhance the wettability of the substrate. This phenomenon is then referred to as electrowetting and has received much attention in the last few decades \cite{McHale2011dielc,Huang2012aphase,Mohseni2006beha,Huo2023thedyna,Zhao2021afini,Hong2008anum,Pillai2021elec}. Roughly speaking, the existing methods for simulation of electrowetting can be classified into two major categories. The first category is the effective contact angle approach \cite{Huang2012aphase,Mohseni2006beha,Huo2023thedyna}. In this approach, the contribution of the electrical field on the contact line is established by the wetting boundary condition characterized by the Young-Lippmann equation. However, since this approach does not consider the governing equations for the electrostatic field, it cannot reproduce the distributions of the electric potential and the free charges in the system. To this end, some researchers have investigated electrowetting using a more realistic mathematical model in which the coupled hydrodynamics-electrostatics equations are included in the system \cite{Zhao2021afini,Hong2008anum,Pillai2021elec}. Following this route, we have noticed that the two fluids used in this approach are typically assumed to be perfect dielectric fluids or follow the leaky dielectric model . Nevertheless, as mentioned in the preceding paragraph, these two models are established by using some assumptions, and thus, they are only applicable to corresponding specific situations. Of late, some researchers have adopted a more general model called the Poisson-Boltzmann equation to investigate electrowetting \cite{Aminfar2009lattice,Goel2023elec,Xu2023elec}. While the Poisson-Boltzmann equation being widely employed in studying ionic distribution in the electric double layer, one point that needs to be mentioned is that the Poisson-Boltzmann equation is derived from the Nernst–Planck equation under some assumptions \cite{Yang2014Acoup}. If these assumptions are not satisfied, such as the cases of the surface being heterogeneously charged \cite{Fu2003ana} or the overlapping electric double layer \cite{Park2007com,Bhattacharyya2009elec}, using the Poisson-Boltzmann equation may lead to incorrect results, and one has to adopt the Nernst–Planck equation in such cases. Many researchers have verified this statement in the single-phase flows \cite{Fu2003ana,Park2007com,Bhattacharyya2009elec}. Likewise, we believe that using the Nernst–Planck model to study the electrowetting multiphase problem should also be more physically plausible than the Poisson-Boltzmann equation. 

The literature survey above indicates that understanding two-phase flows in an electric field requires full consideration of EHD. Unfortunately, since the evolution of the phase interface is inherently nonlinear and strongly coupled with the fluid flow and the electrostatic field, and simultaneously, various variables (such as velocity, density, and electrical properties) vary significantly or even sharply near the phase interface. These characteristics make investigating the interfacial phenomenon for two-phase EHD flow change challenging. Thus, it is desirable to develop numerical methods for such complex problems. Rooted in kinetic theory \cite{Aidun2010lattice,Kruger2017the}, the lattice Boltzmann (LB) method has developed into a powerful and attractive method for simulating ﬂuid ﬂows and complex physical processes. In the past decades, the phase-field-based LB method 
\cite{he1996alatt,liang2014phse} and the pseudopotential LB method 
\cite{Shan1993latt,wang2022therm}, two popular methods for simulating multiphase flow in the LB community, have been widely adopted to simulate two-phase EHD flows. However,  these existing LB methods are almost constructed based on the above-mentioned simplified models, such as the perfect dielectric model \cite{Singh2019elec,Liu2019apha}, the leaky dielectric model \cite{Liu2019apha,Zhang2005A2D}, and the Poisson-Boltzmann model \cite{Aminfar2009lattice,Xu2023elec}, which all share some assumptions to some extent, as mentioned above. Apart from the drawbacks mentioned above, it is also noted that the governing equations for two-phase EHD flow in most previous works are directly established by coupling the hydrodynamic and simplified Maxwell's equations with the interface capturing model. Thus, these LB methods may not be fully consistent with the laws of thermodynamics, which plays a vital role in multiphase flow modeling.

Framed in this general background, the current work tends to propose a thermodynamically consistent phase-field LB method for two-phase EHD flows. To this end, this work first revisits the governing equations for two-phase EHD flows by using the Onsager's variational principle \cite{Qian2006a,Xu2015ava,Xu2017Hydo}. It will be seen later that the thermodynamcially consistent physical model for two-phase EHD flows consist of the Navier-Stokes equations, the Cahn-Hilliard equation coupled with electrical effects, and the full Poisson-Nernst-Planck electrokinetic equations. After that, a general LB method is proposed for simulating two-phase EHD flows based on the resultant physical model. In particular, to simulate the two-phase EHD flows with relatively larger ratios of electric permittivity, which is usually entournted in the electrowetting phenomenon, an improved LB method for electric potential equation is developed. The remainder of the present paper is organized as follows. In Sect. 2, we will derive the governing equations for two-phase EHD flows with Onsager's variational principle. In Sect. 3 and Sect. 4, the corresponding LB method and the boundary conditions are presented. In Sect. 5, we give the iterative algorithm for the current LB method. In Sect. 6, we conduct some numerical simulations to validate the developed LB method. In Sect. 7, the proposed LB method is used to simulate two practical problems. At last, a brief conclusion is drawn in Sect. 8.

\section{The Mathematical model}
In this section, we intend to develop a thermodynamically consistent phase-field model for two-phase EHD flows, in which the electrical and the Cahn-Hilliard equations are coupled to the hydrodynamic equations through extra stress that mimics surface tension and electrical force. In what follows, we first present the generalized governing equations for the two-phase EHD flows. Then, some undetermined terms that appeared in the macroscopic equations are established using Onsager's variational principle to ensure thermodynamic consistency.    

\subsection{The generalized governing equations for the two-phase EHD flows}
\label{sect21}
Assuming the two-fluids are immiscible and incompressible, and neglecting the influence of the gravitational force,  the general formation of the mass and momentum conservation equations in the dimensional form can be expressed as \cite{Castellanos2014elc}
\begin{equation}
	\nabla \cdot \mathbf{u} = 0,
	\label{weq1}
\end{equation}
\begin{equation}
\rho \left( {\frac{{\partial {\bf{u}}}}{{\partial t}} + {\bf{u}} \cdot \nabla {\bf{u}}} \right) = \nabla {\bf{\Pi}} + {\bf{F}},
	\label{eq2}
\end{equation}
where $t$, $\rho$ and $\bm{u}$ are the time, mass density and velocity, respectively. ${\bf{\Pi}}$ is a stress tensor that relates to the hydrodynamic pressure $p$ and the viscous stress tensor $\mu \left[ {\nabla {\bf{u}} + {{\left( {\nabla {\bf{u}}} \right)}^{\rm T}}} \right]$, which can be expressed as  
\begin{equation}
 {\bf{\Pi}}  =  - p{\bf{I}} + \frac{1}{2}\mu \left[ {\nabla {\bf{u}} + {{\left( {\nabla {\bf{u}}} \right)}^{\rm T}}} \right],
\end{equation}
with $\mu$ being the fluid dynamics viscosity. The last term in Eq. (\ref{eq2}), i.e., $\bf{F}$, stands for the external force exerted on the fluids which contains the electrical and interfacial forces and its specific expression will be determined later.   

Apart from the above hydrodynamic equations, considering the essential electrical laws in an EHD flow is another problem that must be specified. Since the current density due to the motion of charge carriers is too small, the influence of the magnetic field can be ignored \cite{Castellanos2014elc}. Consequently, the electric field intensity ${\bf{E}}$ is irrotational, and Gauss's law is restated as \cite{saville1997ehd, Landau1975elec}   
\begin{equation}
\nabla  \cdot \left( {\varepsilon {\bf{E}}} \right) = q, \;\;\;\;\;\;\;\;\; \nabla  \times {\bf{E}} = 0, \;\;\;\;\;\;\;\;\; {\bf{E}} =  - \nabla \varphi ,
\end{equation}
\begin{equation}
 \frac{{\partial q}}{{\partial t}} + {\bf{u}} \cdot \nabla q + \nabla  \cdot {{\bf{J}}_D} = 0,
\label{weq4}
\end{equation}
where $\varepsilon$, $q$ and $\varphi$ denote the permittivity, charge density and potential, respectively. Noting that the last equation in Eq. (\ref{weq4}) is the so-called Nernst–Planck equation with  ${{\bf{J}}_D}$ being the current density due to conduction, which will be derived in what follows. 

Since two-phase EHD flows are typical binary fluid flows, accurately capturing the interface behavior has a significant influence on numerical performance. The current work utilizes the phase-field method \cite{he1996alatt,liang2014phse}, a popular diffuse interface approach for interface tracking, to model the interfacial dynamics. In this setting, the evolution of the order parameter $\phi$ (a parameter introduced to distinguish different phases)  is described by the convective Cahn-Hilliard equation, which can be expressed as 
\begin{equation}
	\frac{{\partial \phi }}{{\partial t}} + \mathbf{u} \cdot \nabla \phi + \nabla  \cdot {\rm{ }}{\mathbf{J}_\phi =0 },
	\label{weq5}
\end{equation} 
where $\mathbf{J}_\phi$ is the undetermined phase flux density.

\subsection{Onsager's variational principle}
As shown in Sec. \ref{sect21}, one can find that there exist some unknown terms (\textit{i.e.}, $\bf{F}$, ${{\bf{J}}_D}$ and ${{\bf{J}}_\phi }$) in the generalized governing equations, which must be established before developing numerical approach. To maintain thermodynamic consistency, we will adopt Onsager's variational principle to determine the governing equations expressed in terms of the above unknowns. To present it more clearly, we first introduce the basic principle of Onsager's variational principle in an isothermal system. We then use it to determine $\bf{F}$, ${{\bf{J}}_D}$ and ${{\bf{J}}_\phi }$ appeared in the governing equations for two-phase EHD flows.

Onsager's variational principle is a variational approach based on the reciprocal relations in linear irreversible thermodynamics \cite{onsager1931recip}. For an open system, the principle states that the evolution equations can be obtained by maximizing the Onsager-Machlup action \cite{Qian2006a,Xu2015ava,Xu2017Hydo},
\begin{equation}
 \mathcal{O} = \dot S + {{\dot S}^*} - {\Phi _S}\left( {\dot \alpha ,\dot \alpha } \right),
\label{weq6}
\end{equation}
where the simple " $ \cdot $" represents the time derivatives, $\dot S$  is the rate of change of the entropy, $ {{\dot S}^*}$ is the rate of entropy given by the system to the environment, ${\Phi _S}\left( {\dot \alpha,\dot \alpha }  \right)$ is the dissipation function defined as the half rate of the entropy production with $\alpha  = \left( {{\alpha _1},{\alpha _2}, \ldots } \right)$ being the set of variables that characterizes the non-equilibrium state of a system. In an isothermal system, it is known that the rate of entropy can be expressed as \cite{Blundell2010con}
\begin{equation}
{{\dot S}^*} =  - \frac{{\dot Q}}{T} =  - \frac{{\dot U}}{T},
\end{equation} 
where $T$ is the temperature, ${\dot Q}$ and ${\dot U}$ are the rate of heat transfer from the environment to the system, and the rate of change of the system energy, and they are equal according to the first thermodynamic law. Note that the rate of change in Helmholtz free-energy and free-energy dissipation function (defined as half the rate of free energy dissipation)  at a constant temperature are respectively written as \cite{Qian2006a,Xu2015ava,Xu2017Hydo},
\begin{equation}
 \mathcal{\dot F} =  - T\left( {S + {{\dot S}^*}} \right),\;\;\;\;\;\;\;\;\;{\Phi _F}\left( {\dot \alpha ,\dot \alpha } \right) = T{\Phi _S}\left( {\dot \alpha ,\dot \alpha } \right).
\label{weq8}
\end{equation}
Submitting Eq. (\ref{weq8}) into Eq. (\ref{weq6}), one can observe that the maximization of the Onsager-Machlup action is equivalent to the minimization of the Rayleighian \cite{Qian2006a,Xu2015ava,Xu2017Hydo}, \textit{i.e.}, 
\begin{equation}
	\mathcal{R} = \mathcal{\dot F}\left( {\dot \alpha ,\dot \alpha } \right) + {\Phi _F}\left( {\dot \alpha ,\dot \alpha } \right).
	\label{weq10}
\end{equation}
Because $\mathcal{\dot F}$  is linear and ${\Phi _F}$ is quadratic in the rates $\dot \alpha$, the above Rayleighian can be  restated as 
\begin{equation}
\mathcal{R} = \sum\limits_i {\frac{{\partial 	\mathcal{F}}}{{\partial {\alpha _i}}}{{\dot \alpha }_i}}  + \frac{1}{2}\sum\limits_{i,j} {{\zeta _{ij}}} {{\dot \alpha }_i}{{\dot \alpha }_j},
\end{equation}
where ${{\zeta _{ij}}}$ is the friction coefficient and it satisfies the reciprocal relation ${\zeta _{ij}} = {\zeta _{ji}}$.  In the end, the kinetic equation is obtained by minimizing $\mathcal{R}$ with regard to the rates,  
\begin{equation}
\sum\limits_j {{\zeta _{ij}}{{\dot \alpha }_j}}  =  - \frac{{\partial \mathcal{F}}}{{\partial {\alpha _i}}}.
\end{equation}

Following the above variational approach, it turns out that if we compute the total Helmholtz free-energy $\mathcal{F}$ and the dissipation function ${\Phi _F}$ for a two-phase EHD flow, then application of Onsager's variational principle could yield the system's time evolution equations. For completeness, the wetting boundary condition that accounts for the contact angle between the phase interface and the solid substrate is also incorporated into the following analysis. In such a case, the Landau free-energy functional in the phase field modelling can be written as \cite{Eck2009on,Landau1975elec} 

\begin{equation}
{{\mathcal{F}}_\phi } = \int_\Omega  {\left( {{\psi_0} + \frac{\kappa }{2}{{\left| {\nabla \phi } \right|}^2}} \right)} d\Omega  + \int_{\partial \Omega } {{\psi _s}} dA,
\end{equation}
where $\Omega$ and ${\partial \Omega }$ are the material volume and its solid boundary, $\psi_0$ is the bulk free-energy density which takes the following double-well form ${\psi_0} = \beta {\phi ^2}{\left( {\phi  - 1} \right)^2}$ \cite{shen2012model}, and this form suggests that the order parameter $\phi$ equal 0 and 1 represent the light and heavy fluids, respectively. ${{\psi _s}}$  is the wall free-energy per unit area at the solid boundary, and $0.5\kappa {\left| {\nabla \phi } \right|^2}$  is excess interfacial energy at the fluid-fluid interface. $\beta$ and $\kappa$ are two coefficients that related to the interface surface tension $\gamma$ and the interface width $W$, and they satisfy $\beta  = {{12\gamma } \mathord{\left/	{\vphantom {{12\gamma } W}} \right.	\kern-\nulldelimiterspace} W},\kappa  = {{3\gamma W} \mathord{\left/{\vphantom {{3\sigma W} 2}} \right. \kern-\nulldelimiterspace} 2}$ \cite{shen2012model,liang2014phse}.
In addition to phase-field free energy, the presence of the electric field also provides a contribution to the total free-energy, which can be expressed as \cite{Landau1975elec} 
\begin{equation}
{{\mathcal{F}}_{\bf{E}} }  = \frac{1}{2}\int_\Omega  {\frac{{{{\left| {\bf{D}} \right|}^2}}}{\varepsilon  (\phi ) }} d\Omega,
\end{equation}
where ${\bf{D}} = \varepsilon (\phi ) {\bf{E}}$ is the electric displacement ﬁeld. Also, the kinetic energy induced by the fluid inertia is given by \cite{Castellanos2014elc}
\begin{equation}
{{\mathcal{F}}_{\bf{u}} }  = \frac{1}{2}\int_\Omega  {\rho {{\left| {\bf{u}} \right|}^2}} d\Omega.
\end{equation}
Additionally, the charge diffusive motion adds to the total free energy a contribution, which can be given by \cite{Eck2009on} 
\begin{equation}
{F_q} = \frac{\lambda }{2}\int_\Omega  {{q^2}} d\Omega,
\end{equation}
in which $\lambda  = {\alpha  \mathord{\left/{\vphantom {\alpha  \sigma }} \right.	\kern-\nulldelimiterspace} \sigma }$ with $\alpha$ and $\sigma$ being the charge diffusion coefficient and electric conductivity, respectively. Framed in the above energy analysis, we can obtain the total Helmholtz free-energy  as
\begin{equation}
	\begin{split}
		{\mathcal{F}} &= {{\mathcal{F}}_\phi } + {{\mathcal{F}}_{\bf{E}}} + {{\mathcal{F}}_{\bf{u}}} + {{\mathcal{F}}_q}\\
		&= \int_\Omega  {\left( {{\psi _0} + \frac{\kappa }{2}{{\left| {\nabla \phi } \right|}^2}} \right)} d\Omega  + \int_{\partial \Omega } {{\psi _s}} dA  + \frac{1}{2}\int_\Omega  {\frac{{{{\left| {\bf{D}} \right|}^2}}}{\varepsilon  (\phi ) }} d\Omega  + \frac{1}{2}\int_\Omega  {\rho {{\left| {\bf{u}} \right|}^2}} d\Omega + \frac{\lambda }{2}\int_\Omega  {{q^2}} d\Omega.
	\end{split}
	\label{weq16}
\end{equation}
Taking the variational operator to the total Helmholtz free-energy, we have  
\begin{equation}
	\begin{split}
		\delta \mathcal{F} &= \int_\Omega  {( - \kappa } {\nabla ^2}\phi  + \frac{{\partial {\psi _0} }}{{\partial \phi }})\delta \phi d\Omega + \int_\Omega  \kappa  \nabla  \cdot (\nabla \phi )\delta \phi d\Omega + \int_{\partial \Omega } {\frac{{\partial {{\psi _s}}}}{{\partial \phi }}} \delta \phi dA - \int_\Omega  {\frac{{\mathop \varepsilon^{\prime} (\phi )}}{{2{\varepsilon ^2}(\phi )}}} |\mathbf{D}{|^2}\delta \phi d\Omega \\
		&= \int_\Omega  {[ - \kappa } {\nabla ^2}\phi  + \frac{{\partial {\psi _0}}}{{\partial \phi }} - \frac{{\mathop \varepsilon^{\prime} (\phi )}}{{2{\varepsilon ^2}(\phi )}}|\mathbf{D}{|^2}]\delta \phi d\Omega + \int_{\partial \Omega } {( - \kappa \mathbf{{n_w}} \cdot \nabla \phi  + \frac{{\partial {\psi _s}}}{{\partial \phi }}} )\delta \phi dA,
	\end{split}
	\label{weq17}
\end{equation}
where the Gauss integral theorem has been adopted, and ${{\bf{n}}_w}$ is the inward unit normal vector of the solid surface. Minimizing $\mathcal{F} $ with respect to the order parameter $\phi$ yields the following equilibrium conditions, 
\begin{equation}
	- \kappa{\nabla ^2}\phi  + \frac{{\partial {\rm{ }}{\psi _0}}}{{\partial \phi }} - \frac{{\mathop \varepsilon^{\prime} (\phi )}}{{2{\varepsilon ^2}(\phi )}}|\bm{D}{|^2}= \hbar  \equiv {\rm{const}} ,\;\;\;\; {\rm{in}} \;\; \Omega,
\end{equation}
\begin{equation}
	- \kappa {{\bf{n}}_w} \cdot \nabla \phi  + \frac{{\partial {\psi _s}}}{{\partial \phi }} = \mathchar'26\mkern-10mu\lambda = 0, \;\;\;\; {\rm{on}} \;\; \partial \Omega,
	\label{weq19}
\end{equation}
where $\hbar$ is the chemical potential in the bulk region, and Eq. (\ref{weq19}) is the so-called wetting boundary condition. Based on the form of Eq. (\ref{weq17}), the rate of change of the total free-energy can be given by 
\begin{equation}
	\dot{\mathcal{F}} = \int_\Omega  {[\hbar \frac{{\partial \phi }}{{\partial t}}]} d\Omega + \int_{\partial \Omega }[ {\mathchar'26\mkern-10mu\lambda} \frac{{\partial \phi }}{{\partial t}}]dA + \int_\Omega  {(\mathbf{E} \cdot \frac{{\partial \mathbf{D}}}{{\partial t}})d\Omega + \int_\Omega  {(\bm{u} \cdot \frac{{\partial \bm{u}}}{{\partial t}})d\Omega + \lambda \int_\Omega  {(q\frac{{\partial q}}{{\partial t}})d\Omega} } } .
	\label{weq21}
\end{equation}
With Eq. (\ref{weq4}) and Eq. (\ref{weq5}), we note that  
\begin{equation}
\int_\Omega  {\left[ {\hbar \left( {\frac{{\partial \phi }}{{\partial t}} + {\bf{u}} \cdot \nabla \phi } \right)} \right]} d\Omega  =  - \int_\Omega  {\hbar \nabla  \cdot {{\bf{J}}_\phi }} d\Omega  = \int_\Omega  {\nabla \hbar  \cdot {{\bf{J}}_\phi }} d\Omega,
\label{weq22}
\end{equation}
\begin{equation}
\int_\Omega  {\left[ {q\left( {\frac{{\partial q}}{{\partial t}} + {\bf{u}} \cdot \nabla q} \right)} \right]} d\Omega  =  - \int_\Omega  {q\nabla  \cdot {{\bf{J}}_D}} d\Omega  = \int_\Omega  {\nabla q \cdot {{\bf{J}}_D}} d\Omega, 	
\end{equation}
where the impermeability conditions ${{\bf{n}}_w} \cdot {{\bf{J}}_\phi } = 0$, ${{\bf{n}}_w} \cdot {{\bf{J}}_D} = 0$ at the solid surface have been utilized. Submitting Eq. (\ref{weq22}) into Eq. (\ref{weq21}), we obtain
\begin{equation}
	\begin{split}
		\dot{\mathcal{F}} &= \int_\Omega  {[\nabla \hbar  \cdot {\mathbf{J}_\phi } - \hbar \mathbf{u} \cdot \nabla \phi ]} d\Omega + \int_{\partial \Omega } [{\mathchar'26\mkern-10mu\lambda} (\mathop \phi \limits^ \cdot   - \mathbf{u}_\tau   \cdot {(\nabla \phi )}_\tau  )]dA + \int_\Omega  {\mathbf{E} \cdot ( - q\mathbf{u} - \mathbf{{J_D}})d\Omega}   \\
		&\qquad+ \int_\Omega  {(\bm{u} \cdot \mathbf{F} - \mathbf{\Pi} :\nabla \mathbf{u})d\Omega + \int_\Omega  [\mathbf{n_w}  \cdot ( {\mathbf{\Pi}} \mathbf{u}_\tau  )]dA + \lambda \int_\Omega {[\nabla q \cdot {\rm{ }}\mathbf{{J_D}} - q\bm{u} \cdot \nabla q]d\Omega} }, 
	\end{split}
	\label{weq14}
\end{equation}
where ${\bf{\tau }}$ represents the direction tangent to the solid surface.

Assuming that the no-slip boundary holds at the solid surface and when the system is away from its equilibrium state, the energy dissipation in the system arises from the diffusive currents ${{{\bf{J}}_\phi }}$, ${{{\bf{J}}_D }}$ in the bulk region and the material time derivative of $\phi$, defined as $\dot \phi  = {{\partial \phi } \mathord{\left/{\vphantom {{\partial \phi } {\partial t}}} \right.\kern-\nulldelimiterspace} {\partial t}} + {\bf{u}} \cdot \nabla \phi$, at the solid surface. Consequently, the dissipation function  ${\Phi _F}\left( {{{\dot \alpha }_i},\dot \alpha } \right)$ for the two-phase EHD flows can be given as follows \cite{Qian2006a,Eck2009on,Xu2015ava}
\begin{equation}
{\Phi _F} = \int_\Omega  {\frac{{{{\left| {\bf{\Pi }} \right|}^2}}}{{2\mu }}} d\Omega  + \int_\Omega  {\frac{{{{\left| {{{\bf{J}}_\phi }} \right|}^2}}}{{2M}}} d\Omega  + \int_\Omega  {\frac{{{{\left| {{{\bf{J}}_D}} \right|}^2}}}{{2\sigma }}} d\Omega  + \int_{\partial \Omega } {\frac{{{{\dot \phi }^2}}}{{2\Gamma }}} dA,
	\label{weq25}
\end{equation}     
where $M$ is the mobility and $\Gamma$ is a phenomenological parameter. This relation states that the shear viscosity in the bulk region (first term), composition diffusion in the bulk region (second and third terms) and composition relaxation at the solid surface (fourth term) all contribute to the changes in the energy dissipation. Apparently, by submitting Eq. (\ref{weq25}) and Eq. (\ref{weq14}) into Eq. (\ref{weq10}), we can immediately  obtain the expression for the Rayleighian as
\begin{equation}
	\begin{split}
		\mathcal{R} &= \int_\Omega  {\left( {{\psi _0} + \frac{\kappa }{2}{{\left| {\nabla \phi } \right|}^2}} \right)} d\Omega  + \int_{\partial \Omega } {{\psi _s}} d\Omega  + \frac{1}{2}\int_\Omega  {\frac{{{{\left| {\bf{D}} \right|}^2}}}{\varepsilon }} d\Omega  + \frac{1}{2}\int_\Omega  {\rho {{\left| {\bf{u}} \right|}^2}} d\Omega + \frac{\lambda }{2}\int_\Omega  {{q^2}} d\Omega\\
		&+ \int_\Omega  {\frac{{{{\left| {\bf{\Pi }} \right|}^2}}}{{2\mu }}} d\Omega  + \int_\Omega  {\frac{{{{\left| {{{\bf{J}}_\phi }} \right|}^2}}}{{2M}}} d\Omega  + \int_\Omega  {\frac{{{{\left| {{{\bf{J}}_D}} \right|}^2}}}{{2\sigma }}} d\Omega  + \int_{\partial \Omega } {\frac{{{{\dot \phi }^2}}}{{2\Gamma }}} dA.
	\end{split}
\end{equation}
Thereafter, minimizing $\mathcal{R}$ with respect to the rates $\bf{u}$, ${{{\bf{J}}_\phi }}$
and ${{\bf{J}}_D}$ under the incompressible condition given by Eq. (\ref{weq1}), we obtain   
\begin{equation}
{\bf{F}} = \hbar \nabla \phi  + q{\bf{E}} - \frac{1}{2}\nabla \left( {\frac{\alpha }{\sigma }{q^2}} \right),
\label{weq26}
\end{equation}  
\begin{equation}
{{\bf{J}}_\phi } =  - M\nabla \hbar ,\;\;\;\;\;{{\bf{J}}_D} = \sigma \left( {{\bf{E}} - \lambda \nabla q} \right).	
\label{weq27}
\end{equation}
By submitting Eq. (\ref{weq26}) and Eq. (\ref{weq27}) into the generalized equations described in Sect. \ref{sect21}, we obtain the system of system equations for the two-phase EHD flows maintaining thermodynamic consistency,
\begin{equation}
\nabla  \cdot {\bf{u}} = 0,
\label{weq29}
\end{equation}
\begin{equation}
\rho \left( {\frac{{\partial {\bf{u}}}}{{\partial t}} + {\bf{u}} \cdot \nabla {\bf{u}}} \right) =  - \nabla \hat p + \mu {\nabla ^2}{\bf{u}} + \hbar \nabla \phi  + q{\bf{E}},
\label{weq30}	
\end{equation}
\begin{equation}
\frac{{\partial \phi }}{{\partial t}} + {\bf{u}} \cdot \nabla \phi  = \nabla  \cdot M\nabla \hbar ,\;\;\;\;\;\hbar  = 4\beta \phi \left( {\phi  - \frac{1}{2}} \right)\left( {\phi  - 1} \right) - \kappa {\nabla ^2}\phi  - \frac{1}{2}\varepsilon '\left( \phi  \right){\left| {\bf{E}} \right|^2},
\label{weq31}	
\end{equation}
\begin{equation}
\nabla  \cdot \left( {\varepsilon {\bf{E}}} \right) = q,\;\;\;\;\; {\rm{ }}{\bf{E}} =  - \nabla \varphi ,
\label{weq32}
\end{equation}
\begin{equation}
\frac{{\partial q}}{{\partial t}} + {\bf{u}} \cdot \nabla q = \nabla  \cdot \left[ {\alpha \nabla q + \sigma \nabla \varphi } \right].
\label{weq33}
\end{equation}

To conclude this section,  certain remarks on the current physical model are summarized below:  

\textit{ Remark I}: Based on the above governing equations, it is noted that the third term in the chemical potential $0.5\varepsilon '\left( \phi  \right){\left| {\bf{E}} \right|^2}$ is associated with the presence of electric field, which suggests that when one adopts phase-field based method to model two-phase EHD flows, the conventional chemical potential (i.e., $4\beta \phi \left( {\phi  - 0.5} \right)\left( {\phi  - 1} \right) - \kappa {\nabla ^2}\phi $ ) used in traditional phase-field method is not applicable theoretically. Similar term was also obtained by Eck et al. \cite{Eck2009on} and Yang et al. \cite{Yang2023phase}, who used the phase-field-based numerical methods to study the problem of electrowetting on dielectric. 

\textit{ Remark II}: Reduction consistency is a crucial property for phase-field modelling \cite{dong2018multi}. This means that for a two-phase system, if one fluid component is absent, then the governing equations for the two-phase system should accurately reduce to those of the corresponding single-phase system. With reduction consistency in mind, the permittivity of $\varepsilon$ in a two-phase EHD flow must adhere to the following constraint: 
\begin{equation}
\varepsilon '\left( \phi  \right)\left| {_{\phi  = 0}} \right. = 0,\;\;\;\;\;\;\varepsilon '\left( \phi  \right)\left| {_{\phi  = 1}} \right. = 0.
\end{equation}
In such a case, accounting for the unequal permittivity of the two fluids, the Hermite interpolation is employed to handle the permittivity across the fluid-fluid interface,
\begin{equation}
\varepsilon \left( \phi  \right) = {\varepsilon _v} + {\phi ^2}\left( {2\phi  - 3} \right)\left( {{\varepsilon _v} - {\varepsilon _l}} \right),
\end{equation}
where the subscripts $v$ and $l$ denote the vapour and liquid phases, respectively. The other physical variables including density $\rho$, dynamic viscosity $\mu$, and conductivity $\sigma$ are all determined using a simple linear function of the order parameter
\begin{equation}
\rho \left( \phi  \right) = \phi \left( {{\rho _l} - {\rho _v}} \right) + {\rho _v},\;\;\;\;\mu \left( \phi  \right) = \phi \left( {{\mu _l} - {\mu _v}} \right) + {\mu _v},\;\;\;\;\sigma \left( \phi  \right) = \phi \left( {{\sigma _l} - {\sigma _v}} \right) + {\sigma _v}.
\label{weq36}
\end{equation}

\section{Lattice Boltzmann method}
Unlike classical computational ﬂuid dynamics approaches based on the discretization of the macroscopic equations, the LB method is based on the lattice Boltzmann equation, which describes the dynamic evolution process of the distribution function in the discrete velocity space \cite{Aidun2010lattice,Kruger2017the}. For two-dimensional cases, the two-dimensional nine-velocity (D2Q9) discrete lattice velocity is commonly adopted \cite{Kruger2017the}, which is given by
\begin{equation}
{{\bf{c}}_i} = \left\{ \begin{array}{l}
		c\left( {0,0} \right),\;\;\;\;\;\;\;\;\;\;\;\;\;\;\;\;\;\;\;\;\;\;\;\;\;\;\;\;\;\;\;\;\;\;\;\;\;\;\;\;\;\;\;\;\;\;\;\;\;\;\;\;\;\;\;\;i = 0,\\
		c\left( {\cos \left[ {{{\left( {i - 1} \right)\pi } \mathord{\left/
						{\vphantom {{\left( {i - 1} \right)\pi } 2}} \right.
						\kern-\nulldelimiterspace} 2}} \right],\sin \left[ {{{\left( {i - 1} \right)\pi } \mathord{\left/
						{\vphantom {{\left( {i - 1} \right)\pi } 2}} \right.
						\kern-\nulldelimiterspace} 2}} \right]} \right),\;\;\;\;\;\;\;\;\;\;\;\;\;\;\;i = 1,2,3,4,\\
		\sqrt 2 c\left( {\cos \left[ {{{\left( {2i - 9} \right)\pi } \mathord{\left/
						{\vphantom {{\left( {2i - 9} \right)\pi } 4}} \right.
						\kern-\nulldelimiterspace} 4}} \right],\sin \left[ {{{\left( {2i - 9} \right)\pi } \mathord{\left/
						{\vphantom {{\left( {2i - 9} \right)\pi } 4}} \right.
						\kern-\nulldelimiterspace} 4}} \right]} \right),\;\;\;\;\;\;i = 5,6,7,8,
	\label{weq37}
	\end{array} \right.
\end{equation}
where $i$ is the lattice direction, $c = {{\Delta x} \mathord{\left/{\vphantom {{\Delta x} {\Delta t}}} \right.\kern-\nulldelimiterspace} {\Delta t}}$ is the lattice speed with ${\Delta x}$ and ${\Delta t}$ being the lattice spacing and time step. In addition, the sound speed $c_s$ in D2Q9 model is expressed as ${c_s} = {c \mathord{\left/{\vphantom {c {\sqrt 3 }}} \right.\kern-\nulldelimiterspace} {\sqrt 3 }}$, and the weight coefficients in different lattice directions are given by ${\omega _0} = {4 \mathord{\left/{\vphantom {4 9}} \right.\kern-\nulldelimiterspace} 9},\;\;{\omega _{1 - 4}} = {1 \mathord{\left/{\vphantom {1 9}} \right.\kern-\nulldelimiterspace} 9},\;\;{\omega _{5 - 8}} = {1 \mathord{\left/	{\vphantom {1 {36}}} \right.
\kern-\nulldelimiterspace} {36}}$.     

\subsection{Lattice Boltzmann method for hydrodynamic equations}
The LB equation for incompressible fluid flow can be written as \cite{liang2014phse} 
\begin{equation}
	\mathop f\nolimits_i ({\bf{x}} + {{\bf{c}}_i} \Delta t,t + \Delta t) = \mathop f\nolimits_i ({\bf{x}},t) - \displaystyle\frac{1}{{\mathop \tau \nolimits_f }}[\mathop f\nolimits_i ({\bf{x}},t) - \mathop f\nolimits_i^{eq} ({\bf{x}},t)] + \Delta t(1 - \frac{1}{{2\mathop \tau \nolimits_f }})\mathop {\bar R}\nolimits_i({\bf{x}},t),
	\label{weq38}
\end{equation}
where $ \mathop f\nolimits_i ({\bf{x}},t) $ denotes the distribution function at position $ {\bf{x}} $ and time $ t $, $ \Delta t $ is the time step, $ \mathop {\bar R}\nolimits_i ({\bf{x}},t) $ is the discrete forcing term defined as
\begin{equation}
	\mathop {\bar R}\nolimits_i (x,t) = \mathop \omega \nolimits_i [\mathbf{u} \cdot \nabla \rho  + \frac{{\mathbf{c}_i  \cdot \mathbf{F}}}{{\mathop c\nolimits_s^2 }} + \frac{{\mathbf{u}\nabla \rho :(\mathbf{c}_i \mathbf{c}_i  - \mathop c\nolimits_s^2 \mathbf{I})}}{{\mathop c\nolimits_s^2 }}].
	\label{weq39}
\end{equation}
$ \mathop f\nolimits_i^{eq} ({\bf{x}},t) $ is the local equilibrium distribution function given by
\begin{equation}
	\mathop f\nolimits_i^{eq} ({\bf{x}},t) = \left\{ {\begin{array}{*{20}{c}}
			{(\mathop \omega \nolimits_0  - 1)\displaystyle\frac{p}{{\mathop c\nolimits_s^2 }} + \rho \mathop s\nolimits_i (\mathbf{u})},&{i = 0},\\
			{\mathop \omega \nolimits_i \displaystyle\frac{p}{{\mathop c\nolimits_s^2 }} + \rho \mathop s\nolimits_i (\mathbf{u})},&{i \ne 0},
	\end{array}} \right.
	\label{weq40}
\end{equation}
with
\begin{equation}
	\mathop s\nolimits_i (\mathbf{u}) = \omega_i [\frac{{\mathbf{c}_i  \cdot \mathbf{u}}}{{\mathop c\nolimits_s^2 }} + \displaystyle\frac{{\mathop {(\mathbf{c}_i  \cdot \mathbf{u})}\nolimits^2 }}{{2\mathop c\nolimits_s^4 }} - \frac{{\mathbf{u} \cdot \mathbf{u}}}{{\mathop {2c}\nolimits_s^2 }}].
	\label{eq29}
\end{equation}
In addition,  the relaxation time ${\tau _f}$ is related to the density and dynamic viscosity and is defined as 
\begin{equation}
\mu  = \rho c_s^2\left( {{\tau _f} - \displaystyle\frac{1}{2}} \right).
\end{equation}
With the distribution function, the hydrodynamic pressure $p$ and velocity $\bf{u}$ are calculated by 
\begin{equation}
	\rho \mathbf{u} = \sum\limits_i {\mathbf{c}_i{f_i}}  + \displaystyle\frac{1}{2}\Delta t\mathbf{F},
	\label{weq43}
\end{equation}
\begin{equation}
	p = \frac{{\mathop c\nolimits_s^2 }}{{1 - \mathop \omega \nolimits_0 }}[\sum\limits_{i \ne 0} {\mathop f\nolimits_i }  + \displaystyle\frac{{\Delta t}}{2}\mathbf{u} \cdot \nabla \rho  + \rho \mathop s\nolimits_0 (\mathbf{u})].
	\label{eq32}
\end{equation}

\subsection{Lattice Boltzmann equation for Cahn-Hilliard equation}
To reduce the spurious velocities arising from the force imbalance at the discrete level, the present work adopts a well-balanced LB method \cite{Ju2024awe} to solve the Cahn-Hilliard equation, and the corresponding evolution equation reads
\begin{equation}
	\mathop g\nolimits_i (\mathbf{x} + \mathbf{c}_i \Delta t,t + \Delta t) = \mathop g\nolimits_i (\mathbf{x},t) - \displaystyle\frac{1}{{_{\mathop \tau \nolimits_g }}}[\mathop g\nolimits_i (\mathbf{x},t) - \mathop g\nolimits_i^{eq} (\mathbf{x},t)] + \Delta t\mathop G\nolimits_i (\mathbf{x},t) + \frac{1}{2}\mathop {\Delta t}\nolimits^2 \mathop \partial \nolimits_t \mathop G\nolimits_i (\mathbf{x},t),
	\label{weq45}
\end{equation}
with  $ \mathop g\nolimits_i^{eq} (\mathbf{x},t) $ being the equilibrium distribution function given by 
\begin{equation}
	g_i^{eq}\left( {{\bf{x}},t} \right) = \left\{ \begin{array}{l}
		\phi  - \left( {1 - {\omega _0}} \right)\hbar ,\;\;i = 0,\\
		{\omega _i}\hbar ,\;\;\;\;\;\;\;\;\;\;\;\;\;\;\;\;\;i \ne 0.
	\end{array} \right.
	\label{weq46}
\end{equation}
The discrete forcing term $ \mathop G\nolimits_i (\mathbf{x},t) $ is expressed as 
\begin{equation}
	\mathop G\nolimits_i (\mathbf{x},t) = \mathop \omega \nolimits_i (\mathbf{u} \cdot \nabla \phi )[ - 1 + \frac{{\mathbf{I}:(\mathbf{c}_i \mathbf{c}_i  - \mathop c\nolimits_s^2 \mathbf{I})}}{{2\mathop c\nolimits_s^2 }}].
	\label{weq47}
\end{equation}
In addition, the mobility $M$ is expressed as a function of relaxation time $\tau_g$ through
\begin{equation}
	M  = \mathop c\nolimits_s^2(\mathop \tau \nolimits_g  - \frac{1}{2})\Delta t,
\end{equation}
and the order parameter in this model is calculated by  
\begin{equation}
	\phi  = \sum\limits_i {\mathop g\nolimits_i }.
	\label{weq49} 
\end{equation}
Note that there some derivatives appeared in the evolution equation, which need to be numerically calculated in the implementation. In such a case, the backward difference scheme is adopted to compute the time derivative, 
\begin{equation}
	\mathop \partial \nolimits_t (\Theta ) = \frac{{\Theta (\mathbf{x},t) - \Theta (\mathbf{x},t - \Delta t)}}{{\Delta t}}.
\end{equation}   
Moreover, to maintain the numerical accuracy of our LB method, we use conventional second-order isotropic discretization schemes to calculate the gradient and Laplace operator terms \cite{Thampi2013iso},  
\begin{equation}
	\nabla \Theta (\mathbf{x}) = \sum\limits_{i \ne 0} {\displaystyle\frac{{\mathop \omega \nolimits_i \mathbf{c}_i \Theta (\mathbf{x} + \mathbf{c}_i \Delta t)}}{{\mathop c\nolimits_s^2 \Delta t}}} ,
\end{equation}
\begin{equation}
	\mathop \nabla \nolimits^2 \Theta (\mathbf{x}) = \sum\limits_{i \ne 0} {\displaystyle\frac{{\mathop {2\omega }\nolimits_i [\Theta (\mathbf{x} + \mathbf{c}_i \Delta t) - \Theta (\mathbf{x})]}}{{\mathop c\nolimits_s^2 \mathop {\Delta t}\nolimits^2 }}} ,
\end{equation}
with $\Theta$ being any physical variable.

\subsection{Lattice Boltzmann equation for electric potential equation}
It is noted that the electric potential is actually governed by the Poisson equation, i.e, 
\begin{equation}
\nabla  \cdot \varepsilon \left( \phi  \right)\nabla \varphi  + q = 0.
\end{equation}
To solve it using the LB method, the model proposed by Chai et al. \cite{Chai2008ano} is typically adopted in the LB community. In such a case,  the relaxation time for electric potential is defined as 
\begin{equation}
\varepsilon \left( \phi  \right) = \chi c_s^2\left( {{\tau _\varphi } - 0.5} \right)\Delta t, 
\end{equation}
in which $\chi$ is an artificial parameter that adjusts the relaxation time. Frankly, this scheme performs well when the diffusion coefficient is uniform or when there are only slight differences in diffusion. However, in EHD flows, the permittivity ratio is always large \cite{Tomar2007two,Zagnoni2009elec}, indicating that the ratio of relaxation times between two fluids is also substantial. Thus,  the conventional scheme may share slow and inefficient solution procedures \cite{Patil2014multi}. Moreover, it is worth noting that while an artificial parameter could adjust the relaxation time, using inappropriate values may result in unstable outcomes \cite{Xiang2012mo}. In this setting, we aim to develop an improved method for solving the electric potential equation with a large permittivity ratio. With the definition of the variable permittivity, the electric potential equation can be rewritten as 
\begin{equation}
	\nabla  \cdot {\varepsilon _v}\nabla \varphi  + \nabla  \cdot \hat \varepsilon \left( \phi  \right)\nabla \varphi  + q = 0,
	\label{weq52}
\end{equation}
where $\hat \varepsilon \left( \phi  \right) = {\phi ^2}\left( {2\phi  - 3} \right)\left( {{\varepsilon _v} - {\varepsilon _l}} \right)$. Based on the above recast equation, the LB equation for electric potential can be expressed as 
\begin{equation}
	{h_i}(\mathbf{x} + {{\bf{c}}_i}\Delta t, t + \Delta t) = {h_i}(\mathbf{x},t) - \frac{1}{{{\tau _h}}}[{h_i}(\mathbf{x},t) - h_i^{eq}(\mathbf{x},t)] - \frac{\hat \varepsilon \left( \phi  \right)}{{\mathop c\nolimits_s^2 }}\frac{{\mathop \omega \nolimits_i \mathbf{c}_i \nabla \varphi}}{{\tau_h }} {\rm{ + }}\Delta t {\varpi _i } q, 
	\label{weq53}
\end{equation}
where $ {h_i}(\mathbf{x},t) $ is the distribution function and $ h_i^{eq}(\mathbf{x},t) $ is the equilibrium distribution function given by
\begin{equation}
	h_i^{eq}(\mathbf{x},t) = \left\{ {\begin{array}{*{20}{c}}
			{(\mathop \omega \nolimits_0  - 1)\varphi (\mathbf{x},t)},&{i = 0},\\
			{\mathop \omega \nolimits_i \varphi (\mathbf{x},t)},&{i = 1, \cdots ,8},
	\end{array}} \right.
	\label{weq54}
\end{equation}
${\varpi _i}$ represents the diffusional-weights given by ${\varpi _0} = 0,{\varpi _{1 - 8}} = {1 \mathord{\left/	{\vphantom {1 8}} \right.\kern-\nulldelimiterspace} 8}$. In addition, the electric potential in the current model can be calculated by
\begin{equation}
	\varphi  = \sum\limits_{i \ne 0} {\frac{1}{{1 - \mathop \omega \nolimits_0 }}} {h_i}.
	\label{weq55}
\end{equation}
Through the Chapman-Enskog analysis shown in the Appendix A, the relaxation time is defined as  $ \mathop \tau \nolimits_h  = 0.5 + {{\mathop \varepsilon \nolimits_v } \mathord{\left/{\vphantom {{\mathop \varepsilon \nolimits_v } {\mathop c\nolimits_s^2 }}} \right.\kern-\nulldelimiterspace} {\mathop c\nolimits_s^2 }}\Delta t $.
In addition, due to the mesoscopic feature of the LB method,  the potential gradient that appeared in the discrete forcing term can be calculated locally by using the following equation   
\begin{equation}
	\nabla \varphi  =  - \frac{{\sum\limits_i {\mathbf{c}_i } \mathop h\nolimits_i }}{{\hat \varepsilon \left( \phi  \right) + \mathop c\nolimits_s^2 \mathop \tau \nolimits_h \Delta t}}.
\end{equation}

\subsection{Lattice Boltzmann equation for Nernst-Planck equation}
Unlike single-phase flows, in a two-phase EHD system, charges are confined around the interface in a thin transition region. Capturing the charge density in such an area with the present diffusive approach requires converting the Nernst-Planck equation into the following equivalent form \cite{Tomar2007two}, 
\begin{equation}
	\frac{{\partial q}}{{\partial t}} + \mathbf{u} \cdot \nabla q = \nabla  \cdot (\alpha\nabla q) + R.
	\label{weq56}
\end{equation}
Here, the forcing term $R$ is defined as 
\begin{equation}
	R =  - \frac{{\sigma q}}{\varepsilon } + \frac{\sigma }{\varepsilon }\nabla \varepsilon  \cdot \mathbf{E} - \nabla \sigma  \cdot \mathbf{E},
	\label{weq57}
\end{equation}
in which $\nabla  \cdot (\varepsilon \mathbf{E}) = q$ has been used.  Inspired by previous works \cite{Wang2021al,Chai2014non}, the evolution equation for charge density can thus be expressed as \cite{Chai2016Amult}
\begin{equation}
	{l_i}(\mathbf{x} + {{\bf{c}}_i}\Delta t, t + \Delta t) = {l_i}(\mathbf{x},t) - \frac{1}{{{\tau _l}}}[{l_i}(\mathbf{x},t) - l_i^{eq}(\mathbf{x},t)] + \Delta t\mathop S\nolimits_i (\mathbf{x},t) + \Delta t\mathop T\nolimits_i (\mathbf{x},t),
	\label{weq58}
\end{equation}
where $S_i$ and $T_i$ are two discrete forcing terms and they are given by
\begin{equation}
	\mathop S\nolimits_i (\mathbf{x},t) = (1 - \frac{1}{{2{\tau _l}}})\mathop \omega \nolimits_i R, 
	\label{weq59}
\end{equation}
\begin{equation}
	\mathop T\nolimits_i (\mathbf{x},t) = (1 - \frac{1}{{2{\tau _l}}})\frac{{\mathop \omega \nolimits_i {{\bf{c}}_i} \cdot \mathop \partial \nolimits_t (q{\bf{u}})}}{{\mathop c\nolimits_s^2 }}.
	\label{weq60}
\end{equation}
In addition, the local equilibrium function is defined as
\begin{equation}
	l_i^{eq}(\mathbf{x},t) = \mathop \omega \nolimits_i q(1 + \frac{{{{\bf{c}}_i}  \cdot {\bf{u}}}}{{\mathop c\nolimits_s^2 }}).
	\label{weq61}
\end{equation}
The charge density is determined by 
\begin{equation}
	q = \sum\limits_i {{l_i}}  + \frac{1}{2}\Delta tR,
	\label{weq62}
\end{equation}
and the relaxation time $\tau_l$ is evaluated by $\mathop \tau \nolimits_l  = 0.5 + {\alpha \mathord{\left/
		{\vphantom {D {\mathop c\nolimits_s^2 }}} \right.
		\kern-\nulldelimiterspace} {\mathop c\nolimits_s^2 }}\Delta t$.

\section{Wetting boundary condition     }
\label{sect4}
When two-phase fluids encounter a solid substrate, the substrate’s wettability significantly influences fluid interface dynamics. Therefore, it is imperative to establish a robust wetting boundary condition that considers the contact angle between the phase interface and the solid surface. This section will present the details of implementing the wetting boundary condition with the current LB method.   

As shown in Eq. (\ref{weq19}), the wetting boundary condition can be defined once the wall free-energy ${\psi _s}$ is determined. In this work, we are using the cubic wall free-energy approach \cite{Khatavkar2007cap}, which only takes into account the interaction at the three-phase junction and ignores the interactions between the solid and bulk phases, and its express can be defined as 
\begin{equation}
{\psi _s} = \frac{{{b_1}}}{2}{\phi ^2} - \frac{{{b_1}}}{3}{\phi ^3},
\label{weq63}
\end{equation}
where $b_1$ is a model parameter. Additionally, following the route provided in Ref.\cite{Liang2019la}, it is noted that ${\psi _s}$ in the bulk region must meet with   
\begin{equation}
\frac{{d{\psi _s}}}{{d\phi }} =  \pm \sqrt {2\kappa {\psi _0}}. 
\label{weq64}
\end{equation}
Combing Eq. (\ref{weq63}) and Eq. (\ref{weq64}), one can find that ${\psi _s}$ shares two stable solutions, i.e., ${\psi _{s1}} = 0$ and ${\psi _{s2}} = 1$. Thus, the surface tensions for gas-solid and liquid-solid can be given by 
\begin{equation}
{\gamma _{sg}} = \frac{{{b_1}}}{2}\psi _{s1}^2 - \frac{{{b_1}}}{3}\psi _{s1}^3 + \int_0^{{\psi _{s1}}} {\sqrt {2\kappa {\psi _0}} } d\phi  = 0,
\label{weq65}
\end{equation}

\begin{equation}
{\gamma _{sl}} = \frac{{{b_1}}}{6}\psi _{s2}^2 - \frac{{{b_1}}}{3}\psi _{s2}^3 + \int_1^{{\psi _{s2}}} {\sqrt {2\kappa {\psi _0}} } d\phi  = \frac{{{b_1}}}{6}.
\label{weq66}
\end{equation}
It is known that the local static contact angle ${\theta _Y}$ on a chemically homogeneous wall satisﬁes the Young's equation
\begin{equation}
\cos {\theta _Y} = \frac{{{\gamma _{sg}} - {\gamma _{sl}}}}{\gamma }.
\label{weq67}
\end{equation}
Based on Eqs.  (\ref{weq19}), (\ref{weq65}) and (\ref{weq66}), we can ultimately obtain the wetting boundary condition as
\begin{equation}
	\mathbf{n}_w  \cdot \nabla \phi  =  - \sqrt {\frac{{2\beta }}{\kappa }} \cos {\theta _Y} (\phi  - \mathop \phi \nolimits^2 ).
	\label{weq68}
\end{equation}

It is worth noting that the wetting boundary condition mentioned above is established without considering the influence of the electric field. Since Lippmann's pioneering work on electro-capillarity \cite{Lippmann1875rela}, it has been discovered that the divergent electric force leads to a significant deformation of the fluid interface and causes a large curvature in the vicinity of the contact line. In such situations, the apparent contact angle ${\theta _B}$ can be described by the Lippmann equation \cite{Lippmann1875rela},
\begin{equation}
\cos {\theta _B} = \cos {\theta _Y} + \frac{{{\varepsilon _s}{\varphi ^2}}}{{2\gamma d}},
\label{weq69}
\end{equation}
where $\varphi $ is the applied voltage, $\varepsilon _s$ and $d$ are the permittivity and thickness of the dielectric substrate, respectively. In the literature, the second term on the right-hand side of Eq. (\ref{weq69}) is also referred to as the electrowetting number. This number is used to measure the relative strength of the electric force in contrast to the surface tension at the fluid surface, and it can be rewritten as
\begin{equation}
{{{\varepsilon _s}{\varphi ^2}} \mathord{\left/
			{\vphantom {{{\varepsilon _s}{\varphi ^2}} {\left[ {2\gamma d} \right]}}} \right.
			\kern-\nulldelimiterspace} {\left[ {2\gamma d} \right]}} = {\eta  \mathord{\left/
			{\vphantom {\eta  d}} \right.
			\kern-\nulldelimiterspace} d},
\end{equation}
in which $\eta  = {{0.5{\varepsilon _s}{\varphi ^2}} \mathord{\left/{\vphantom {{0.5{\varepsilon _s}{\varphi ^2}} \gamma }} \right.\kern-\nulldelimiterspace} \gamma }$. Comparing Eq. (\ref{weq67}) and Eq. (\ref{weq69}), it is found that the electrowetting boundary condition can be established by using an effective surface tension for liquid-solid, which is given by 
\begin{equation}
{{\hat \gamma }_{sl}} = {\gamma _{sl}} - \frac{{{\varepsilon _s}{\varphi ^2}}}{{2\gamma d}}.
\end{equation}
In summary, the electrowetting boundary condition is still implemented with Eq. (\ref{weq68}) expect that the previous surface tension for liquid-solid ${\gamma _{sl}}$ is replaced by ${{\hat \gamma }_{sl}}$.

\section{Implementation scheme of the proposed LB method}
Due to the nonlinear coupling characteristic of governing equations, we use an iterative scheme outlined in Algorithm \ref{algorith1} to implement the current LB method. The key algorithm of this approach involves a cyclic sequence of substeps, where each cycle corresponds to one time step. In each iteration, the macroscopic variables in the system are updated by solving the LB equation for every physical field. It is worth noting that because the electric potential equation is governed by a steady elliptic equation, an inner loop is required to obtain the convergent solution for the electric potential at each time step.

\begin{breakablealgorithm}
	\label{algorith1}
	\caption{Implementation scheme of the proposed LB method }
	\begin{algorithmic}
		\STATE $\#$ 1. At the beginning ($n=0$), the fluid variables ($\phi$, $\rho$, ${\bf{u}}$, $\varphi$, $q$) and the distribution functions ($g_i^{0} , h_i^{0}, l_i^{0}, f_i^{0}$) at each grid point $\mathbf{x}_i$ are initialized 
		
		\STATE$\#$ 2. Solve the Chan-Hilliard equation Eq. (\ref{weq31}) using Eq. (\ref{weq45})
		\FORALL{$\mathbf{x}_i$}
		\STATE $\mathop g\nolimits_i^{{\rm{n + 1}}}  \leftarrow \mathop g\nolimits_i^n  - \displaystyle\frac{1}{{\mathop \tau \nolimits_g }}[\mathop g\nolimits_i^n  - \mathop g\nolimits_i^{eq,\;n} ] + \Delta t\mathop G\nolimits_i^n  + \frac{1}{2}\mathop {\Delta t}\nolimits^2 \mathop \partial \nolimits_t \mathop G\nolimits_i^n $,
		\STATE $\#$ where $g_i^{eq,\; n}$ and $G_i^{n}$ are calculated with Eq. (\ref{weq46}) and Eq. (\ref{weq47}),
		\STATE $\phi^{n+1} \leftarrow \sum\limits_i {g_i^{n+1} }$,
		\STATE $\rho^{n+1} \left( \phi  \right) \leftarrow \phi^{n+1} \left( {{\rho _l} - {\rho _v}} \right) + {\rho _v}$.
		\ENDFOR
		
		$\#$ 3. Solve the electric potential equation Eq. (\ref{weq52}) using Eq. (\ref{weq53}) \\
		\FORALL{$\mathbf{x}_i$}
		\STATE $\mathop h\nolimits_i^{{\rm{n + 1}}}  \leftarrow \mathop h\nolimits_i^n  - \displaystyle\frac{1}{{\mathop \tau \nolimits_h }}[\mathop h\nolimits_i^n  - \mathop h\nolimits_i^{eq,\;n} ] - \frac{{\hat \varepsilon^{n} (\phi )}}{{\mathop c\nolimits_s^2 }}\frac{{\mathop \omega \nolimits_i \mathbf{c}_i \nabla \phi^{n} }}{{\mathop \tau \nolimits_h }} + \Delta t\mathop \varpi \nolimits_i q^{n}$,
		\STATE $\#$ where $h_i^{eq,\; n}$ and $\varphi$ are calculated utilizing Eq. (\ref{weq54}) and Eq. (\ref{weq55}),
		\STATE $\varphi^{n+1} \leftarrow \sum\limits_{i \ne 0} {\displaystyle\frac{1}{{1 - \mathop \omega \nolimits_0 }}} {h_i^{n+1}}$,
		\STATE repeat until $\sqrt {\displaystyle\frac{{\sum\limits_i {\mathop {[\mathop \varphi \nolimits_i^{n + 1}  - \mathop \varphi \nolimits_i^n ]}\nolimits^2 } }}{{\sum\limits_i {\mathop {[\mathop \varphi \nolimits_i^n ]}\nolimits^2 } }}}  < \mathop {10}\nolimits^{ - 5}$.
		\ENDFOR
		
		$\#$ 4. Solve the Nernst-Planck equation Eq. (\ref{weq56}) using Eq. (\ref{weq58})\ \\
		\FORALL{$\mathbf{x}_i$}
		\STATE $\mathop l\nolimits_i^{{\rm{n + 1}}}  \leftarrow \mathop l\nolimits_i^n  - \displaystyle\frac{1}{{\mathop \tau \nolimits_l }}[\mathop l\nolimits_i^n  - \mathop l\nolimits_i^{eq,\; n} ] + \Delta t\mathop S\nolimits_i^n  + \Delta t\mathop T\nolimits_i^n$,
		\STATE $\#$ where $S_i^{n}$, $T_i^{n}$ and $l_i^{eq,\; n}$ are calculated using Eqs. (\ref{weq59}), (\ref{weq60}) and  (\ref{weq61}), respectively,
		\STATE $q^{n+1} \leftarrow \sum\limits_i {{l_i^{n+1}}}  + \displaystyle\frac{1}{2}\Delta tR^{n}$.
		\ENDFOR
		
		$\#$ 5. Solve the hydrodynamic equations Eq. (\ref{weq29}) and Eq. (\ref{weq30}) using  Eq. (\ref{weq43})\ \\
		\FORALL{$\mathbf{x}_i$ and $t$}
		\STATE $\mathop f\nolimits_i^{{\rm{n + 1}}}  \leftarrow \mathop f\nolimits_i^n  - \displaystyle\frac{1}{{\mathop \tau \nolimits_f }}[\mathop f\nolimits_i^n  - \mathop f\nolimits_i^{eq,\; n} ] + \Delta t(1 - \frac{1}{{2\mathop \tau \nolimits_f }})\mathop {\bar R}\nolimits_i^n$,
		\STATE $\#$ where ${\bar R}_i^{n}$ and $f_i^{eq,\; n}$ are calculated using Eq. (\ref{weq39}) and Eq. (\ref{weq40}),
		\STATE $\rho \mathbf{u}^{n+1} \leftarrow \sum\limits_i {\mathbf{c}_i{f_i^{n+1}}}  + \displaystyle\frac{1}{2}\Delta t\mathbf{F}^{n}$.
		\ENDFOR
		
		$\#$ 6. Advance the time step ($n+1 \leftarrow n$) and repeat step 2 to 5 until the end time or steady state is reached
	\end{algorithmic}
\end{breakablealgorithm}

\section{Numerical Validation}
In this section, we will validate the proposed thermodynamically consistent phase-field LB method by simulating the EHD flows with two superimposed planar fluids \cite{Herrera2011ac}, the charge relaxation of a Gaussian bell \cite{Herrera2011ac} and the equilibrium interface profile of electrowetting on dielectric. Our primary focus will be on its performance in predicting the electric potential, charge density, and contact angle of a charged droplet.

\subsection{EHD flows with two superimposed planar ﬂuids}

\begin{figure}[H]
	\centering
	\includegraphics[width=0.35\textwidth]{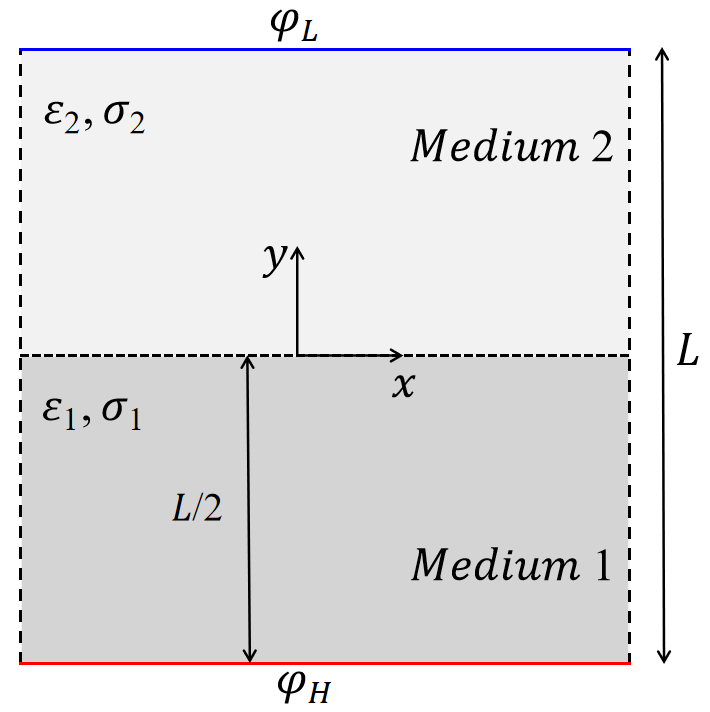}				
	\caption{Schematic diagram of EHD flows with two superimposed planar ﬂuids.}
	\label{wfig1}
\end{figure}

\begin{figure}[H]
	\centering
	\subfigure[]{
		\label{wfig2a} 
		\includegraphics[width=0.45\textwidth]{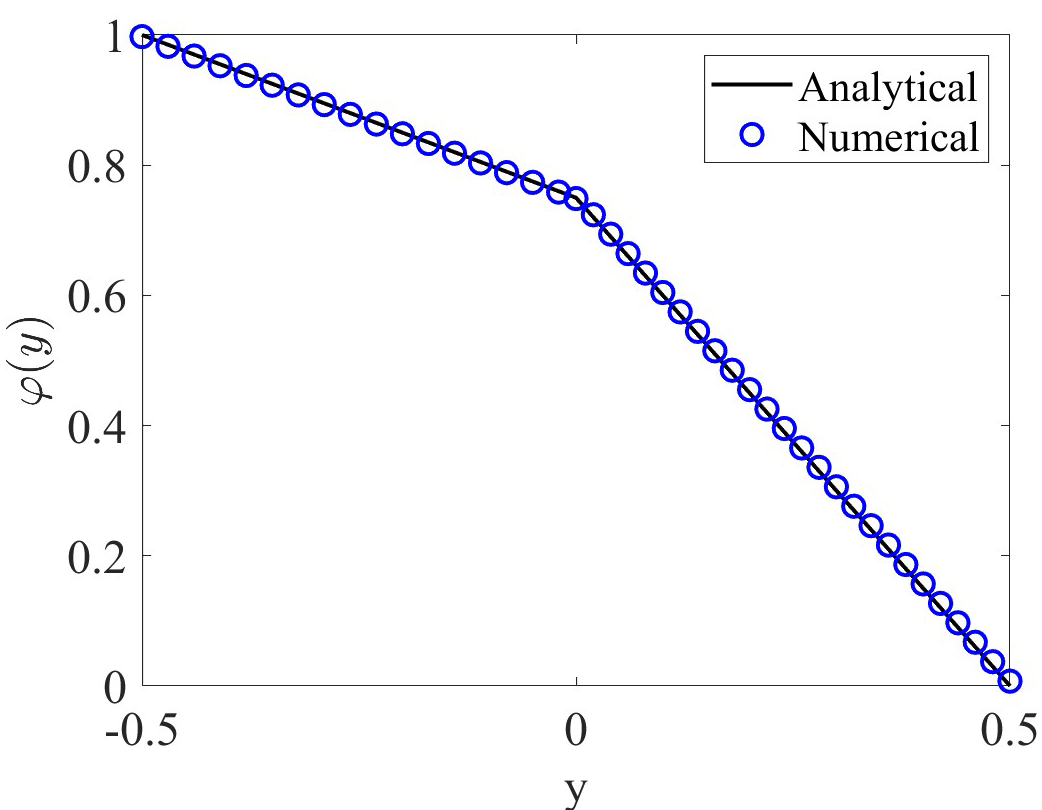}}
	\subfigure[]{
		\label{wfig2b} 
		\includegraphics[width=0.46\textwidth]{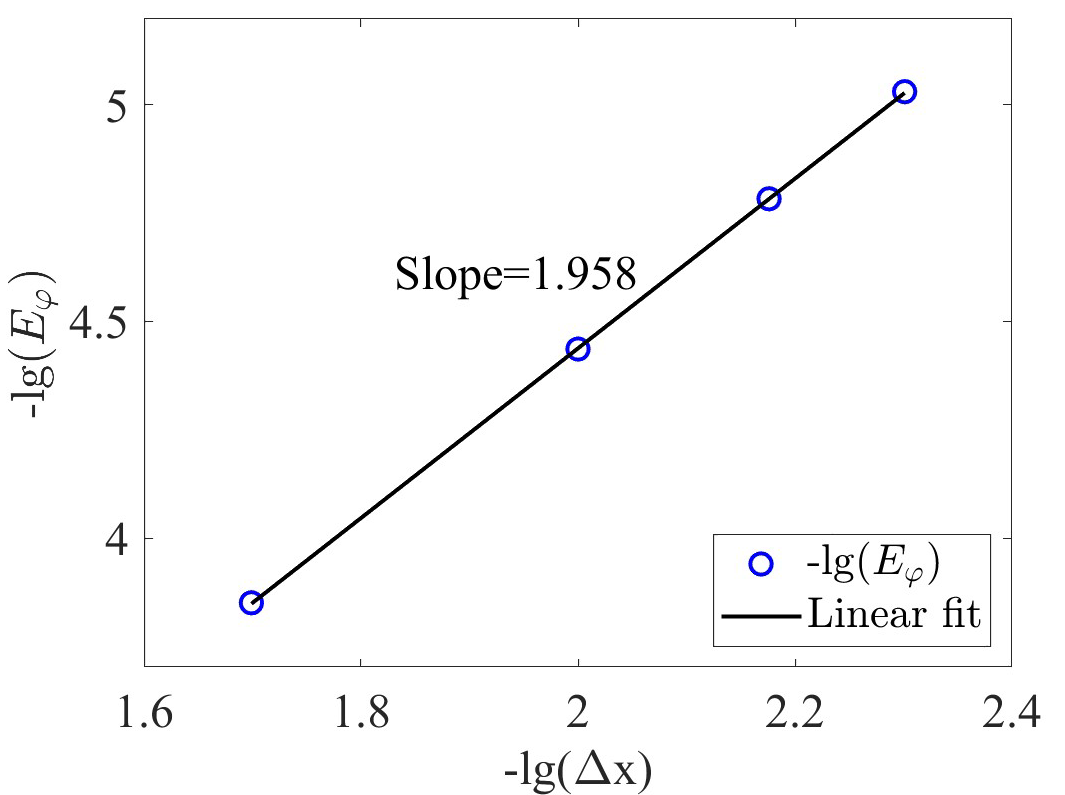}}
	
	\caption{(a) Comparison of the potential between analytical solutions given by Eq. (\ref{weq72}) and numerical results for the EHD flows with two superimposed planar ﬂuids; (b) Relative error of the electric potential versus the mesh size, in which the ratios of the droplet's conductivity and permittivity to surrounding fluids are set to be 3.0 and 2.0, respectively.} 
	\label{wfig2}
\end{figure}

We first validate the proposed model by simulating the EHD flows with two superimposed planar fluids \cite{Herrera2011ac}. The configuration is shown in Fig. \ref{wfig1}, in which a square enclosure with width being $L$ is filled with two conducting fluids, and the height of these two fluids are both set to be $L/2$. Note that the electric properties of these two fluids are homogeneous, and the potential at the upper wall and bottom wall are given by ${\varphi _L}$ and ${\varphi _H}$ (${\varphi _H} > {\varphi _L}$), respectively. According to the work of López-Herrera et al. \cite{Herrera2011ac}, the analytical solution for the electric potential in each medium can be given by 
\begin{equation}
	\mathop \varphi \nolimits_1^{exact} (y) = \frac{{ - 2y + \hat R}}{{1 + \hat R}},  \;\;\;\;
	\mathop \varphi \nolimits_2^{exact} (y) = \frac{{ - 2y + 1}}{{1 + \hat R}}\hat R,
	\label{weq72}
\end{equation}
where $\hat R$ denotes the ratio of conductivities between two fluids i.e., ${{\hat R = {\sigma _l}} \mathord{\left/
{\vphantom {{R = {\sigma _l}} {{\sigma _v}}}} \right.\kern-\nulldelimiterspace} {{\sigma _v}}}$. Not that the electric properties used in our simulations are the same as those adopted by López-Herrera et al. \cite{Herrera2011ac}. Fig. \ref{wfig2a} gives the profiles of $\varphi$ along the $y$ coordinate, in which the mesh is set to be $100 \times 100$. It is shown that our numerical data give reasonably accurate results. To evaluate the convergence rate of the proposed LB model, the relative errors under different grid resolutions are further calculated, which is defined as 

\begin{equation}
	E_{\varphi} = \sqrt {\frac{{\sum\limits_i {\mathop {(\mathop \varphi \nolimits_{numerical} (\mathbf{x}_i) - \mathop \varphi \nolimits_{analytical} (\mathbf{x}_i ))}\nolimits^2 } }}{{\sum\limits_i {\mathop {\mathop \varphi \nolimits_{analytical} (\mathbf{x}_i )}\nolimits^2 } }}}, 
\end{equation}
where the subscripts "analytical" and "numerical" represent the analytical solution and numerical result, respectively, and the summation is over the entire domain. As seen from Fig. \ref{wfig2b}, the current LB scheme shares a second order accuracy in space. The above numerical results demonstrate the feasibility of the present treatment for electrostatic field, and verifies that the electric potential equation can be solved properly by the current proposed model. 

\subsection{Charge diffusion of a Gaussian bell}

\begin{figure}[H]
	\centering
	\subfigure[]{
		\label{wfig3a} 
		\includegraphics[width=0.45\textwidth]{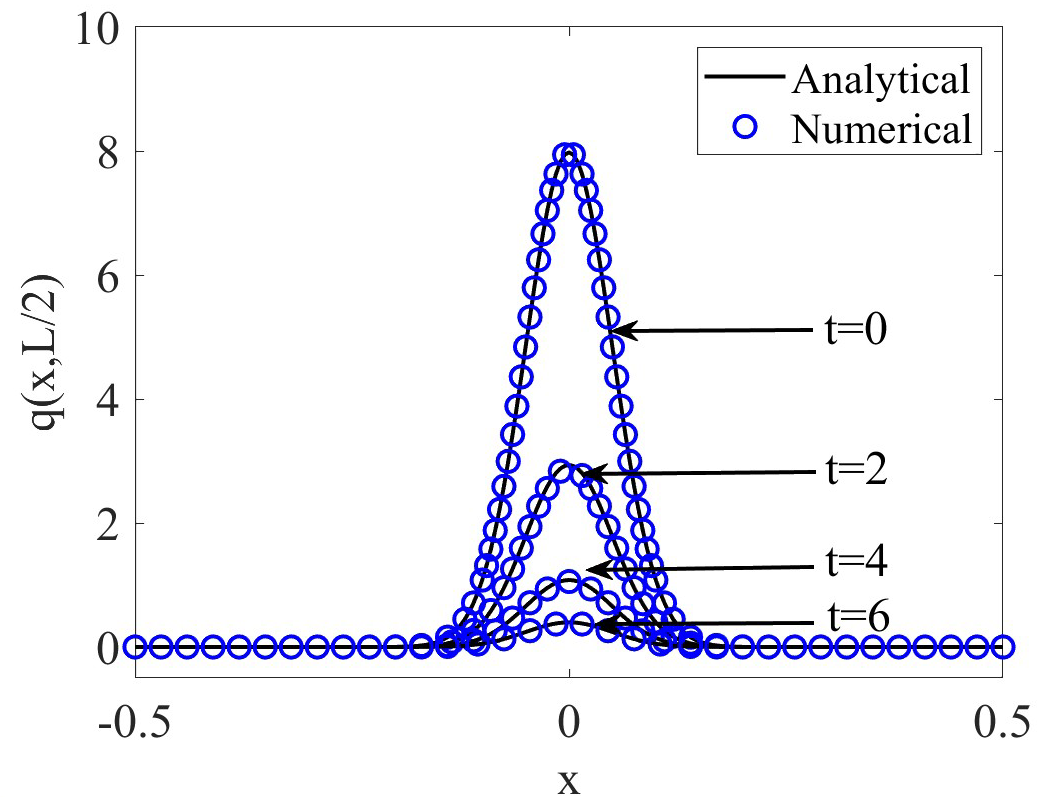}}
	\subfigure[]{
		\label{wfig3b} 
		\includegraphics[width=0.45\textwidth]{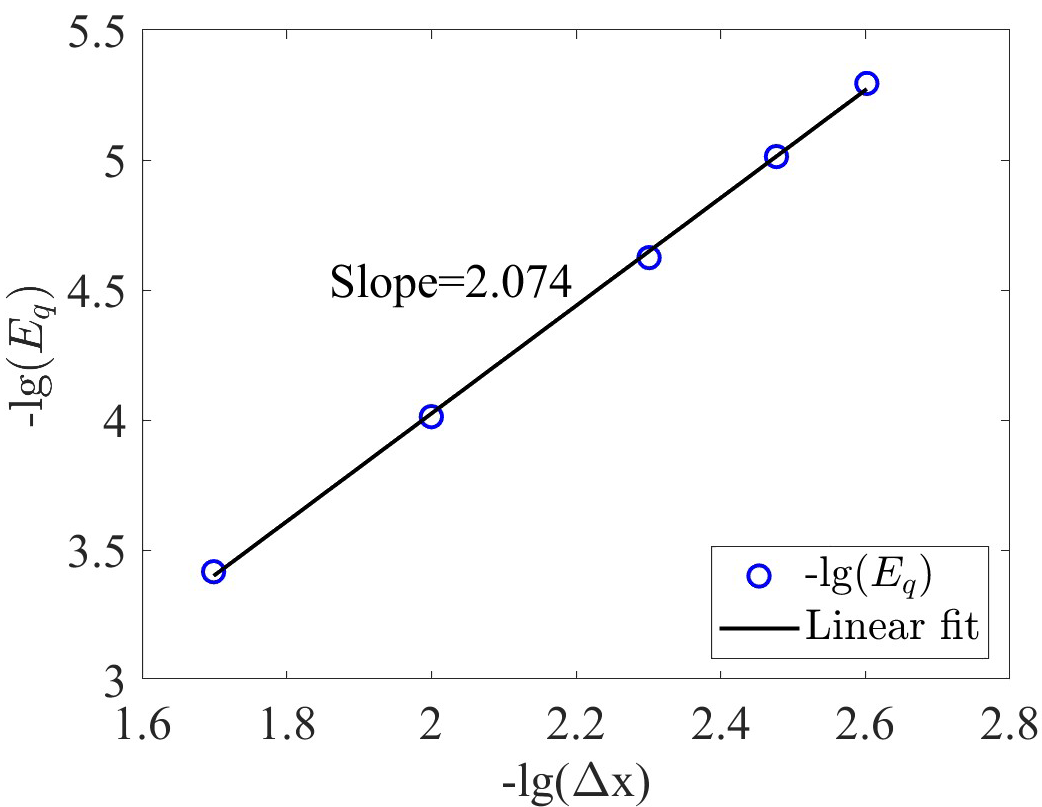}}			
	
	\caption{(a) Comparison of the charge density at diffrtent time between the analytical solution given by Eq. (\ref{weq78}) and the numerical results for charge diffusion of a Gaussian bell; (b) relative error of the charge density versus the mesh size, in which the simulated parameters are chosen as $\sigma=1.0$, $\varepsilon =2.0$, $a=0.05$ and $L=1.0$.}
	\label{wfig3}
\end{figure}

In order to further verify the proposed model, another test problem discussed is the time diffusion of a charge density distribution \cite{Herrera2011ac}. In this problem, a square enclosure with width of $L$ is occupied by a single-phase fluid, and the electric potential at the four boundaries is set to be 0.0. The initial shape of the charge density obeys a Gaussian bell defined as
\begin{equation}
q\left( {x,y,t = 0} \right) = \frac{{\exp \left( {\frac{{ - {\iota ^2}}}{{2{a^2}}}} \right)}}{{a\sqrt {2\pi } }},
\end{equation}
where ${\iota ^2} = {x^2} + {y^2}$ and $a$ is a free parameter determining the width and height of the bell. If the domain boundaries are sufficiently far from the concentrated charge bump, i.e., $a \ll L$, the analytical solution for the charge density can be expressed as     
\begin{equation}
q\left( {x,y,t} \right) = \frac{{\exp \left( {\frac{{ - {\iota ^2}}}{{2{a^2}}} + \frac{{\sigma t}}{\varepsilon }} \right)}}{{a\sqrt {2\pi } }}.
\label{weq78}
\end{equation}

We now perform some simulations under different times with the conductivity, permittivity and free parameter being $\sigma=1.0$, $\varepsilon=2.0$, and $a=0.05$, respectively. In addition, the width of the square enclosure is set to be $L=1.0$, which is large enough to derive the analytical solution. Fig. \ref{wfig3a} presents the comparison between the numerical results and analytical solution for the charge density along the horizontal center line, where the lattice size is $200 \times 200$. Good consistence can be observed, which verifies the applicability of the current model for Nernst-Planck equation. Further, the convergence rate of the present model is investigated as well. To this end, the relative error defined by Eq. (\ref{weq72})  is calculated with grid resolution varying from 50 to 400 at $t=2$. As shown in Fig. \ref{wfig3b}, one can clearly seen that our model shares a second-order convergence rate, which is consistent with the theoretic analysis.

\subsection{Equilibrium interface profiles}

The above two simulations are conducted without considering the influence of the wetting behaviour. In this example, we intend to test the model's capability in predicting the equilibrium interface profiles and compare them with the analytical solutions predicted by the Lippmann equation. In the simulation, the lattice size of the computation domain is set to be $200 \times 100$, which is fine enough to give the grid independence results. Initially, a conductive droplet with a radius of $r=25$ is located on the bottom wall, and the area surrounding the droplet is filled with an insulating gas. The order profile function is defined as
\begin{equation}
	\phi {\rm{ = 0}}{\rm{.5 + 0}}{\rm{.5tanh\{ }}\frac{{{\rm{2}}{\rm{.0}} \times [{\rm{r - }}\sqrt {\mathop {(x - 200)}\nolimits^2  + \mathop {(y - 1)}\nolimits^2 } ]}}{W}\},
\end{equation} 
in which the interface thickness $W$ is fixed at four lattice units in our simulations. The density ratio  (${{{\rho _l}} \mathord{\left/	{\vphantom {{{\rho _l}} {{\rho _v}}}} \right.	\kern-\nulldelimiterspace} {{\rho _v}}}$), the kinetic viscosity ratio (${{{\mu _l}} \mathord{\left/{\vphantom {{{\mu _l}} {{\mu _v}}}} \right.	\kern-\nulldelimiterspace} {{\mu _v}}}$), and the permittivity ratio are set to be 100, 100, and 81, respectively. The top wall is kept at a low electric potential $\varphi_L$, and the bottom wall is held at a high electric potential $\varphi_H$. In addition, apart from the periodic boundary condition used in the horizontal direction and the no-slip bounce-back boundary condition imposed at the rigid walls \cite{Kruger2017the}, the wetting boundary condition described by Eq. (\ref{weq68}) should also be adopted for the bottom wall. Moreover, to examine the effect of an electric field on wetting behaviour,  it is essential to form a steady droplet pattern before applying the electric field. Thus, in our simulation, we first run the code without an electric field until it reaches the prescribed Young's contact angle. Following this, the droplet will be relaxed by incorporating the effect of an electric field and ultimately stabilize into an equilibrium pattern.

\begin{figure}[H]
	\centering
	\subfigure[]{
		\label{wfig4a} 
		\includegraphics[width=0.45\textwidth]{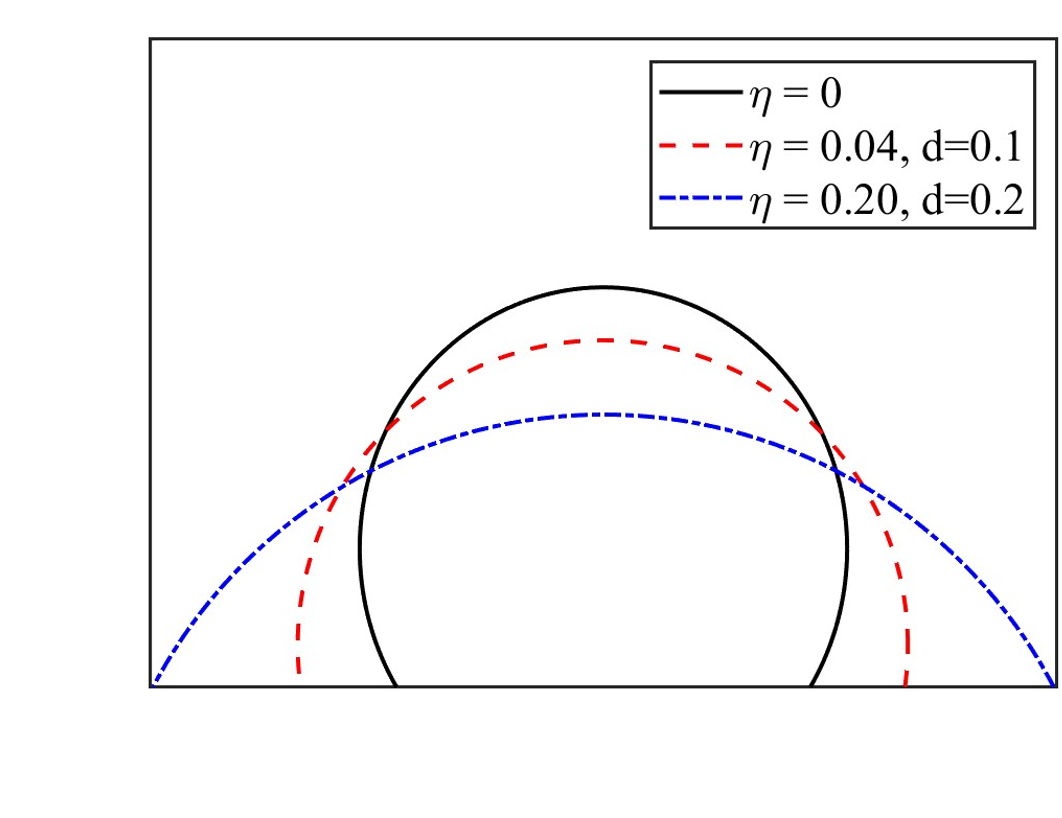}}		
	\subfigure[]{
		\label{wfig4b} 
		\includegraphics[width=0.45\textwidth]{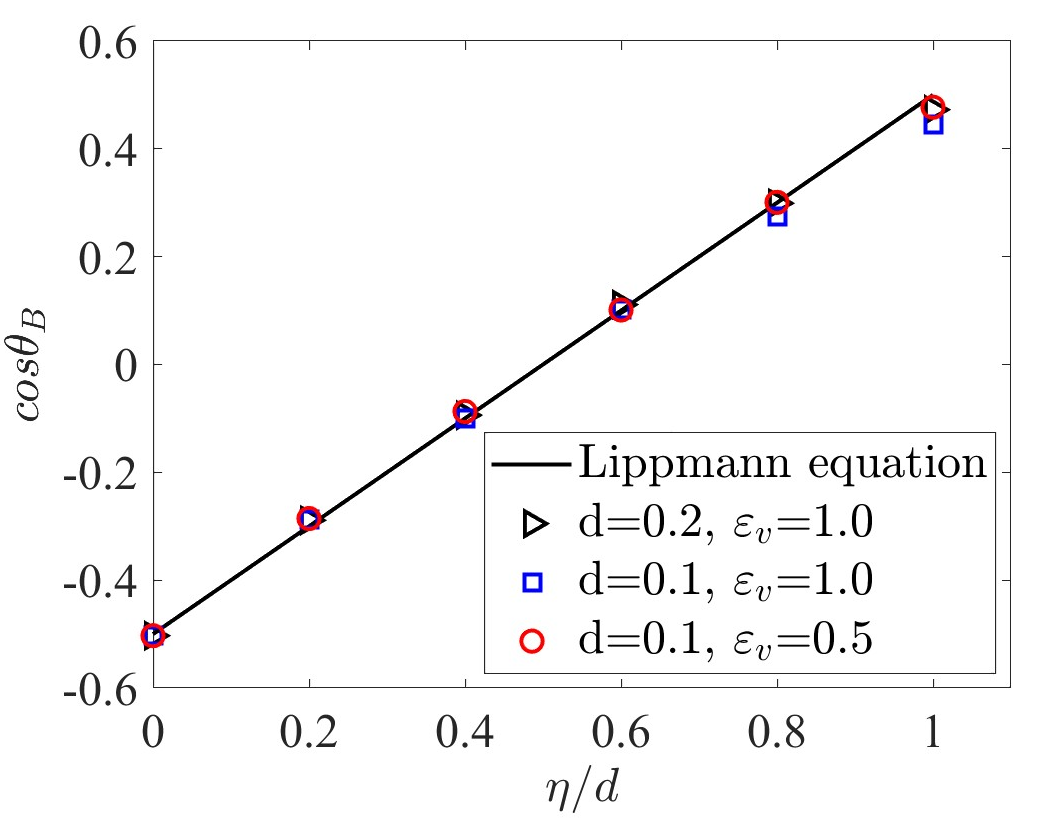}}
	
	\caption{(a) Equilibrium profile of the droplet interface for different values of the dimensionless parameters $\eta$ and $d$; (b) Variation of $cos{\theta}_{B}$ versus $\eta/d$ and the comparison with predictions of the Young-Lippmann equation.} 
	\label{wfig4}
\end{figure}

In Fig. \ref{wfig4a}, the graph illustrates the change in the equilibrium interface profile of the droplet for different $\eta$ and $d$ with ${{\varepsilon _v}}=1.0$. It is evident that when a voltage is applied between the conductive droplet and the solid surface coated with a dielectric material, the change of interfacial energy leads to a decrease in the apparent contact angle compared to the case in the absence of the electric field. To quantify the numerical results, we present the variation of $\cos {\theta _{B}}$ for different  $d$ and ${{\varepsilon _v}}$. It is observed that $\cos {\theta _{B}}$  increases linearly with a slope close to 1, and the permittivity of the fluid outside the droplet has an insignificant influence on the interface profile. The current results align closely with the analytical solutions provided by the Lippmann equation (Fig. \ref{wfig4b}).

\section{Applications}
In this section, the proposed method is used to simulate two practical problems: droplet deformation in an electric field and droplet detachment in reversed electrowetting. The first problem demonstrates our model's ability to capture the fluid-fluid interface dynamics under an electric field. On the other hand, the second problem shows that the proposed method can replicate the motion of the contact line of a charged droplet. The simulated results will be compared against theoretical, experimental, and previous numerical data to validate their accuracy.  

\subsection{Droplet deformation in an electric field}

\begin{figure}[H]
	\centering
	\includegraphics[width=0.35\textwidth]{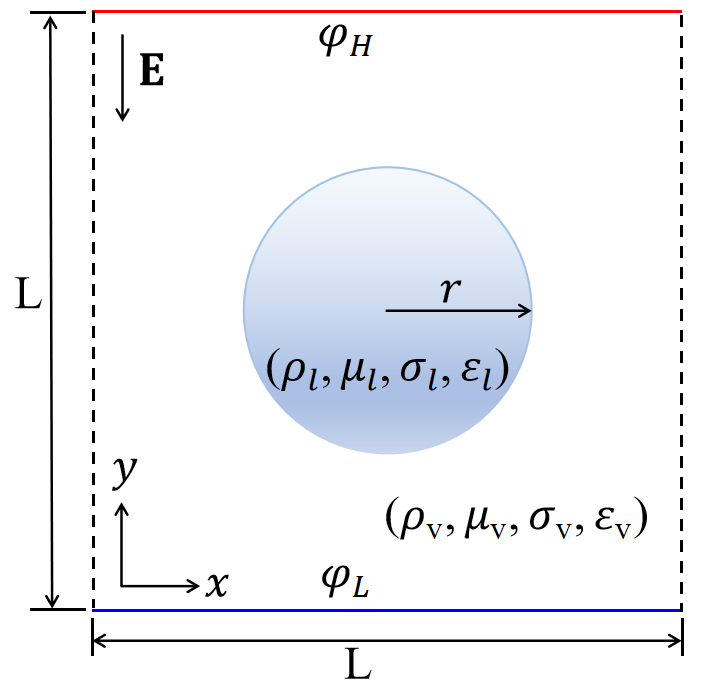}				
	\caption{The configuration of a droplet suspended in another leaky dielectric fluid under a uniform electric field. The subscripts $ l $ and $ v $ represents the internal and external fluids, respectively.}
	\label{wfig5}
\end{figure}

The deformation of a droplet in the presence of an externally applied electric field is a critical scientific concern in the EHD community. It has been found that when a uniform electric field is applied, the droplet can deform either a prolate shape (aligned with the electric field) or an oblate shape (perpendicular to the electric field), which is a result of the combined effects of conductivity and permittivity ratios. To show that the current model is also capable of handling such a complex problem, some simulations are carried out and our numerical results are compared with the theoretical solutions and the available numerical data.   

The configuration is depicted in Fig. \ref{wfig5}. Initially, the droplet suspended in another leaky dielectric fluid is a circle droplet with a radius of $r$, and it is placed in the middle of a square cavity. To induce the droplet deformation,  a uniform electric field $\bf{E}$ is applied in the vertical direction, and the electric potentials at the top wall and the bottom wall are set to be ${\varphi _H}$ and ${\varphi _L}$ (${\varphi _H} > {\varphi _L}$), respectively. In this setting, the droplet deformation in the presence of an electric field can be characterized utilizing a parameter $\hat D$ given by \cite{Taylor1964Dis}
\begin{equation}
	\hat D = \frac{{\hat L - \hat H}}{{\hat L + \hat H}},
\end{equation}
in which $\hat L$ denotes the length of the droplet in the direction parallel to the electric field, $\hat H$ is the length perpendicular to the electric field. Based on this equation, it is clear that the droplet deforms into a prolate (oblate) shape when the deformation factor $\hat D$ is larger (smaller) than zero. Assuming that the two fluids are extremely viscous and conducting, Taylor's small deformation theory predicts the deformation factor $\hat D$ at steady state as \cite{Taylor1964Dis} 
\begin{equation}
	\hat D = \frac{9}{{16}}\frac{{\mathop {Ca}\nolimits_E }}{{\mathop {(2 + \hat R)}\nolimits^2 }}[\mathop {\hat R}\nolimits^2  + 1 - 2\hat S + \frac{3}{5}(\hat R - \hat S)\frac{{2 + 3\hat B}}{{1 + \hat B}}],
	\label{eq53}
\end{equation}
where $\hat R$, $\hat S$ and $\hat B$ denotes the ratios of the droplet's conductivity, permittivity, and viscosity to the surrounding fluid, respectively. The electric capillary number $\mathop {Ca}\nolimits_E  = {{\mathop \varepsilon \nolimits_v \mathop E\nolimits_0^2 r} \mathord{\left/{\vphantom {{\mathop \varepsilon \nolimits_v \mathop E\nolimits_0^2 r} \gamma }} \right.\kern-\nulldelimiterspace} \gamma }$ is the ratio of electric stress ($u_e$) to capillary stress ($u_c$). It should be noted that Taylor's small deformation theory is constructed by ignoring the surface charge convection, which suggests that the multiphase interface is charged instantaneously. However, the above assumption cannot hold well when the charge relaxation time is comparable to the flow and capillary time scales. In such a case, Feng considered the influence of the surface charge convection and proposed an improved formation for the deformation factor $\hat D$ \cite{Feng1999Elec}
\begin{equation}
 	\hat D = \frac{{\mathop {Ca}\nolimits_E }}{{3\mathop {(1 + 2\hat R)}\nolimits^2 }}[\mathop {\hat R}\nolimits^2  + \hat R + 1 - 3\hat S].
 	\label{eq54}
\end{equation}
Note that when the surface charge convection is incorporated, apart from the above mentioned electric capillary number $Ca_E$, we also need to consider another three dimensionless numbers \cite{Luo2020Num} , i.e.,   
\begin{equation}
	{\mathop{\rm Re}\nolimits}  = \frac{{\mathop r\nolimits^2 \mathop \rho \nolimits_v \mathop \varepsilon \nolimits_v \mathop E\nolimits_0^2 }}{{\mathop \mu \nolimits_v^2 }}, \qquad \mathop {{\mathop{\rm Re}\nolimits} }\nolimits_E  = \frac{{\mathop \varepsilon \nolimits_v^2 \mathop E\nolimits_0^2 }}{{\mathop \mu \nolimits_v \mathop \sigma \nolimits_v }},\qquad \alpha  = \frac{{\omega \mathop k\nolimits_B T\mathop \mu \nolimits_v }}{{\mathop \varepsilon \nolimits_v \mathop r\nolimits^2 \mathop E\nolimits_0^2 }},
\end{equation}
where $Re$ is the Reynolds number defined as the ratio of electric force to viscous force, $Re_E$ is the electric Reynolds number used to quantify the surface charge convection, $\alpha$ is the charge diffusion coefficient with $k_B$, $\omega$ and $T$ being the Boltzmann constant, charge mobility and fluid temperature, respectively.

In the simulation, the square width $L$ is set to be $L=8r$, and the lattice size of the computation domain is fixed at $200 \times 200$. The periodic boundary condition is adopted in the horizontal direction, while on the top and bottom plates, the boundary conditions are realized using the half-way bounce-back scheme \cite{Kruger2017the}. The density ratio is set to be 2.0 in order to avoid remarkable droplet migration driven by buoyancy \cite{Feng1996acom}. The other parameters used in the phase-field simulation are chosen as $ \gamma  = 0.001 $ (surface tension), $ W = 5.0 $ (interface width), $ M = 0.1 $ (mobility). Additionally, in order to compare the current numerical results with the theoretical and numerical solutions obtained from the leaky dielectric model, the value of $Re_{E}$ should be sufficiently small, i.e., ${{\mathop{\rm Re}\nolimits} _E} \ll 1$, such that the ohmic conduction mechanism is dominant over the charge convection mechanism in the system. Also, according to the experiment, the Reynolds number is in the order of ${\rm O}\left( 1 \right)$, and the charge diffusion coefficient is about  ${10^{ - 4}}$. With the above analysis, unless otherwise stated, the following simulations in this subsection are all performed at $Re_{E}=10^{-3}$, $Re=1.0$, $\alpha=10^{-4}$ and $Ca_{E}=0.2$.    

\begin{figure}[H]
	\centering
	\subfigure[]{
		\label{wfig6a} 
		\includegraphics[width=0.3\textwidth]{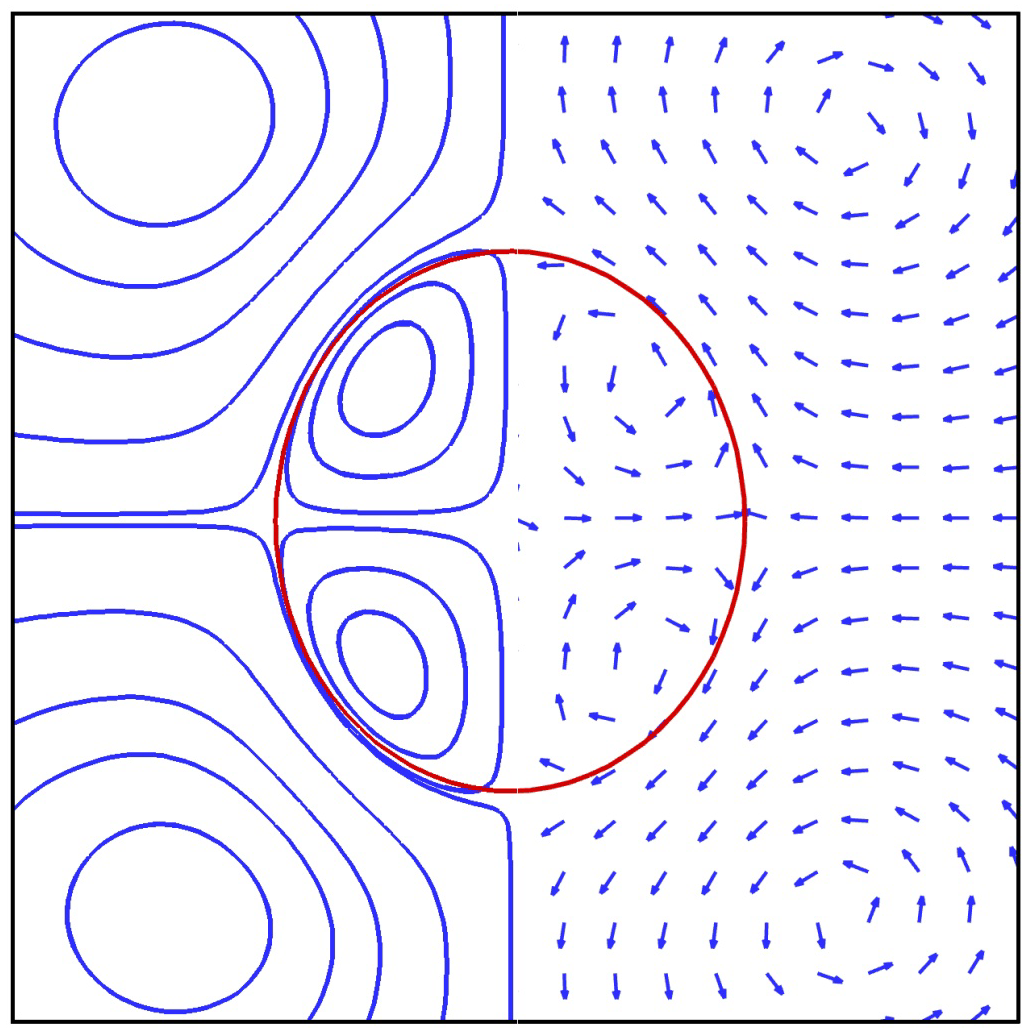}}		
	\subfigure[]{
		\label{wfig6b} 
		\includegraphics[width=0.3\textwidth]{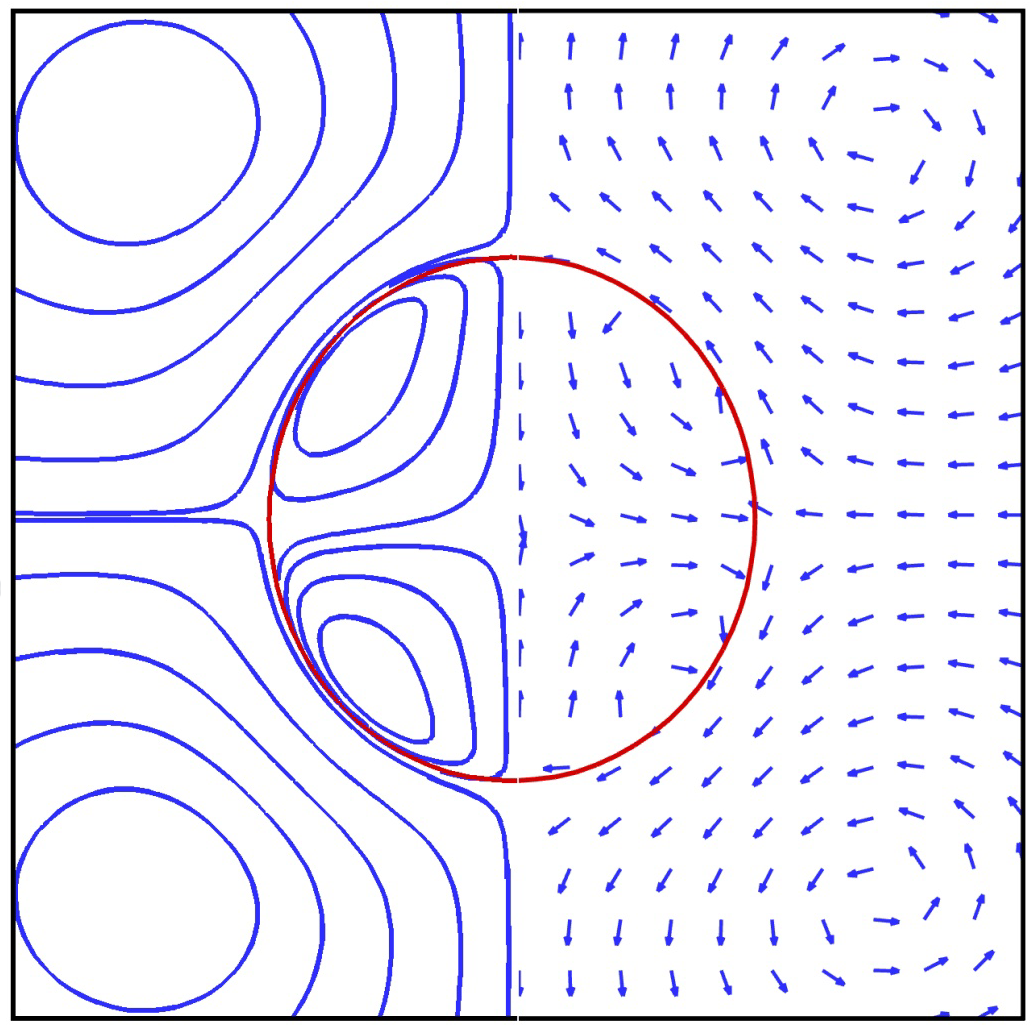}}
	\subfigure[]{
		\label{wfig6c} 
		\includegraphics[width=0.3\textwidth]{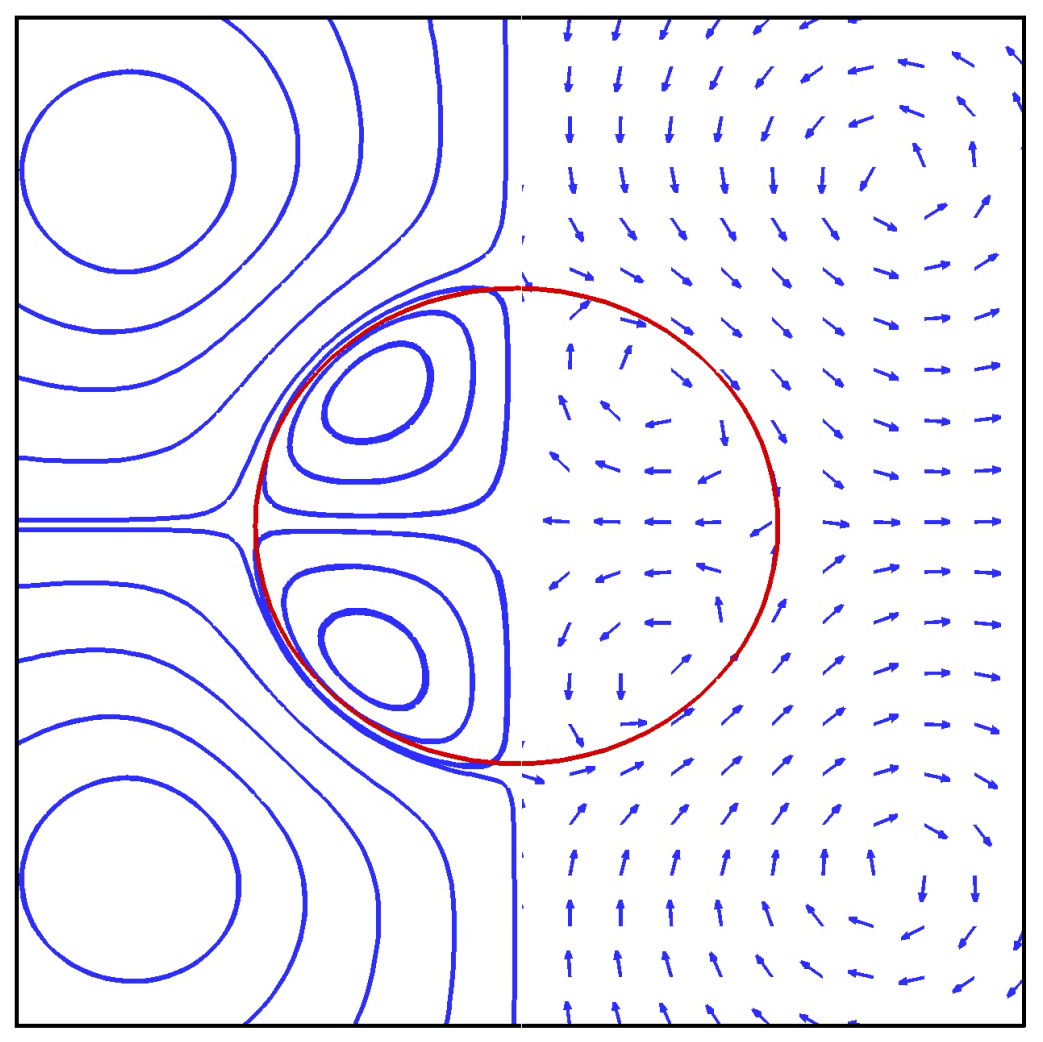}}		
	
	\caption{Stream lines (left half), velocity vectors (right half) and the profile of the droplet (red lines) for three representative steady-state droplets: (a) prolate deformation obtained at $(\hat R, \hat S)=(5, 0.5)$; (b) slight prolate deformation at $(\hat R, \hat S)=(5, 5)$; (c) oblate deformation at $(\hat R, \hat S)=(5, 15)$, , in which the simulated parameters are chosen as $Ca_E=0.2$, ${{\mathop \rho \nolimits_l } \mathord{\left/{\vphantom {{\mathop \rho \nolimits_l } {\mathop \rho \nolimits_v }}} \right.\kern-\nulldelimiterspace} {\mathop \rho \nolimits_v }} = 2.0$, ${{\mathop \mu \nolimits_l } \mathord{\left/{\vphantom {{\mathop \mu \nolimits_l } {\mathop \mu \nolimits_v }}} \right.\kern-\nulldelimiterspace} {\mathop \mu \nolimits_v }} = 1.0$.}
	\label{wfig6}
\end{figure}

\begin{figure}[H]
	\centering
	\subfigure[]{
		\label{wfig7a} 
		\includegraphics[width=0.3\textwidth]{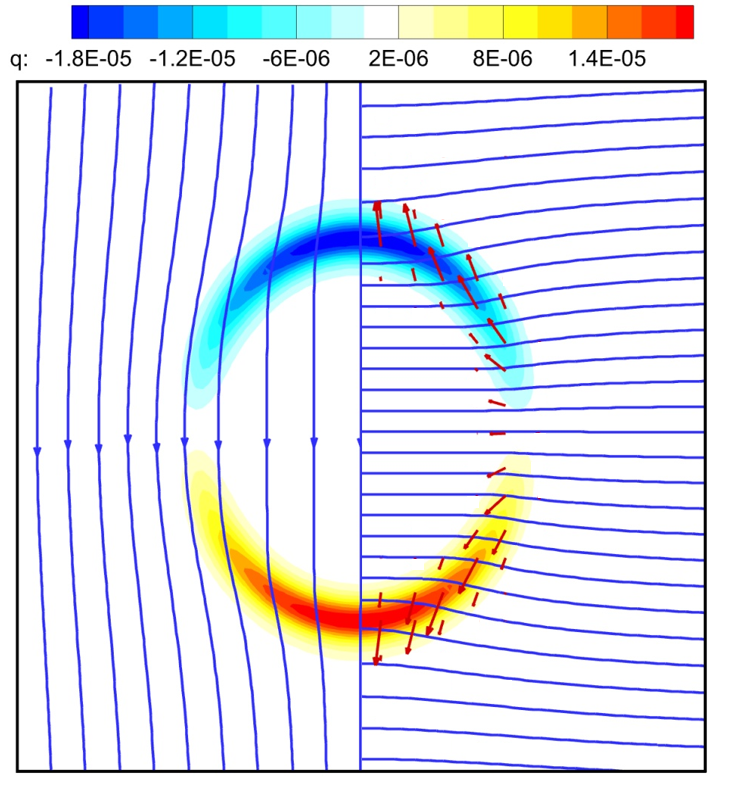}}		
	\subfigure[]{
		\label{wfig7b} 
		\includegraphics[width=0.3\textwidth]{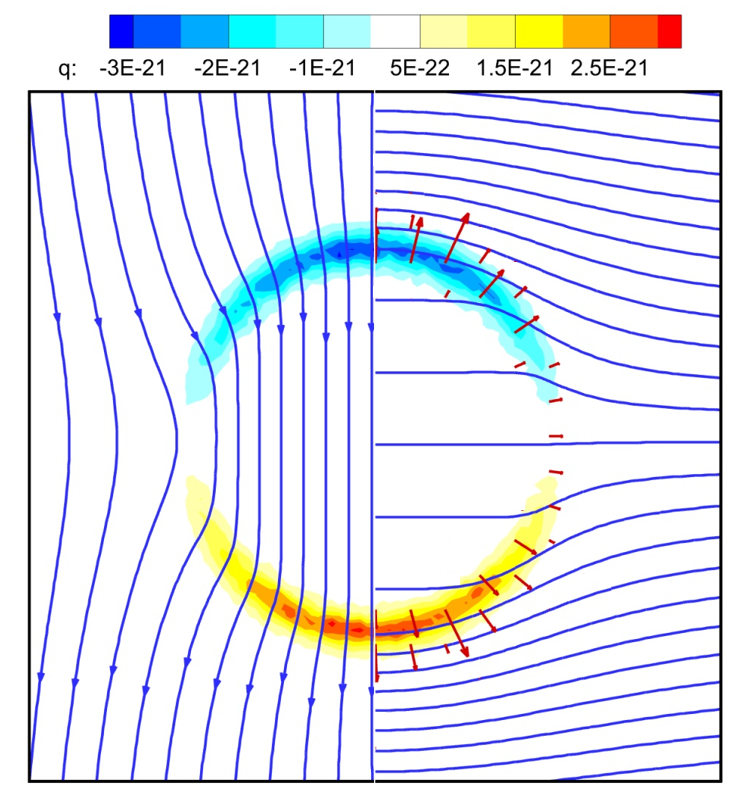}}
	\subfigure[]{
		\label{wfig7c} 
		\includegraphics[width=0.3\textwidth]{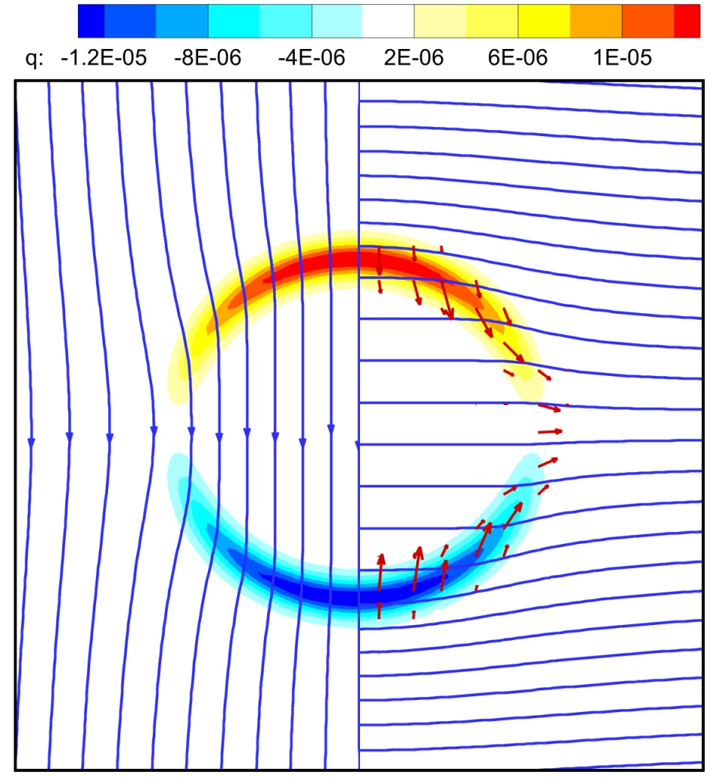}}		
	
	\caption{Variation in electric field lines (left half), electric potential lines (right half), charge distribution (colored parts) and the electric field force (red arrows) for three different values of permittivity ratios: (a) $\hat S=0.5$; (b) $\hat S=5.0$; (c) $\hat S=15.0$, in which the simulated parameters are chosen as $Ca_E=0.2$, ${{\mathop \rho \nolimits_l } \mathord{\left/{\vphantom {{\mathop \rho \nolimits_l } {\mathop \rho \nolimits_v }}} \right.\kern-\nulldelimiterspace} {\mathop \rho \nolimits_v }} = 2.0$, ${{\mathop \mu \nolimits_l } \mathord{\left/{\vphantom {{\mathop \mu \nolimits_l } {\mathop \mu \nolimits_v }}} \right.\kern-\nulldelimiterspace} {\mathop \mu \nolimits_v }} = 1.0$, $\hat R=5.0$.} 
	\label{wfig7}
\end{figure}

Fig. \ref{wfig6} presents the streamlines and velocity vectors for different permittivity ratios $\hat S$ with conductivity ratio  $\hat R$ being fixed at 5.0. As shown in Fig. \ref{wfig6a}, it is seen that for the case of $(\hat R, \hat S)= (5,0.5)$, the droplet deforms into a prolate shape with four symmetrical and counter-rotating vortices formed inside and outside of the droplet, and the vector diagram in the ﬁrst quadrant of the droplet displays that the circulation direction is from the equator to the poles along the droplet surface. When $(\hat R, \hat S)= (5,5)$, the deformation of the droplet is insignificant, and flow patterns are similar to those observed at $(\hat R, \hat S)= (5,0.5)$ (see Fig. \ref{wfig6b}). However, when the ratio of the electric permittivities ($\hat S=15.0$) is larger than the ratio of the electrical conductivities ($\hat R= 5$),  an oblate droplet with reversed flow direction is obtained. To have a better understanding on the influence of the electric field, Fig. \ref{wfig7} illustrates the distributions of the electric potential, charge density, electric field lines, as well as the electric force direction for different permittivity ratios. It is evident that due to the discontinuity in the permittivity at the interface, the electric potential and electric field lines are distorted in this region. Specifically, a fluid with high permittivity polarizes more in response to an applied electric field, resulting in denser potential lines inside the droplet for a relatively smaller permittivity ratio. Additionally, it is noted that for all cases, the maximum charge density appears at the poles while it is negligible at the equator. In particular, we also found that the position of the positive and negative charges is related to the droplet's shape, which directly leads to a difference in the direction of the corresponding electric field force.

\begin{table}[H]
	\centering
	\caption{Quantitative comparison of the results from the present work with theoretical solutions and existing data.}
	\label{tab:tb1}
	\setlength{\tabcolsep}{5mm}{
		\begin{tabular}{llllllll}
			\hline
			\hline
			\multirow{2}*{$\hat R$}  & \multirow{2}*{$\hat S$} & \multirow{2}*{$ Ca_E $} & \multicolumn{4}{c}{Deformation factor ($\hat D$)} \\
			\cline{4-7}
			{} & {} & {} & Present & Eq.(\ref{eq53}) & Eq.(\ref{eq54}) & Ref. \cite{Luo2020Num} \\
			\hline
			5     & 5    & 0.2 		  &  0.03503   &  0.03670      &  0.02960    &  0.03150      \\
			5     & 60   & 0.2        & -0.26304   & -0.40520      & -0.27590    & -0.27510      \\
			1     & 2    & 0.2        & -0.04102   & -0.04380      & -0.05000    & -0.05390      \\
			1.75  & 3.5  & 0.1        & -0.01968   & -0.02230      & -0.02070    & -0.02000      \\     
			3.25  & 3.5  & 0.1        &  0.00848   &  0.00850      &  0.00800    &  0.00900      \\
			4.75  & 3.5  & 0.1        &  0.02236   &  0.02280      &  0.01800    &  0.02090      \\
			\hline   
			\hline
		\end{tabular}
	}
\end{table}

\begin{figure}[H]
	\centering
	\subfigure[]{
		\label{wfig8a} 
		\includegraphics[width=0.45\textwidth]{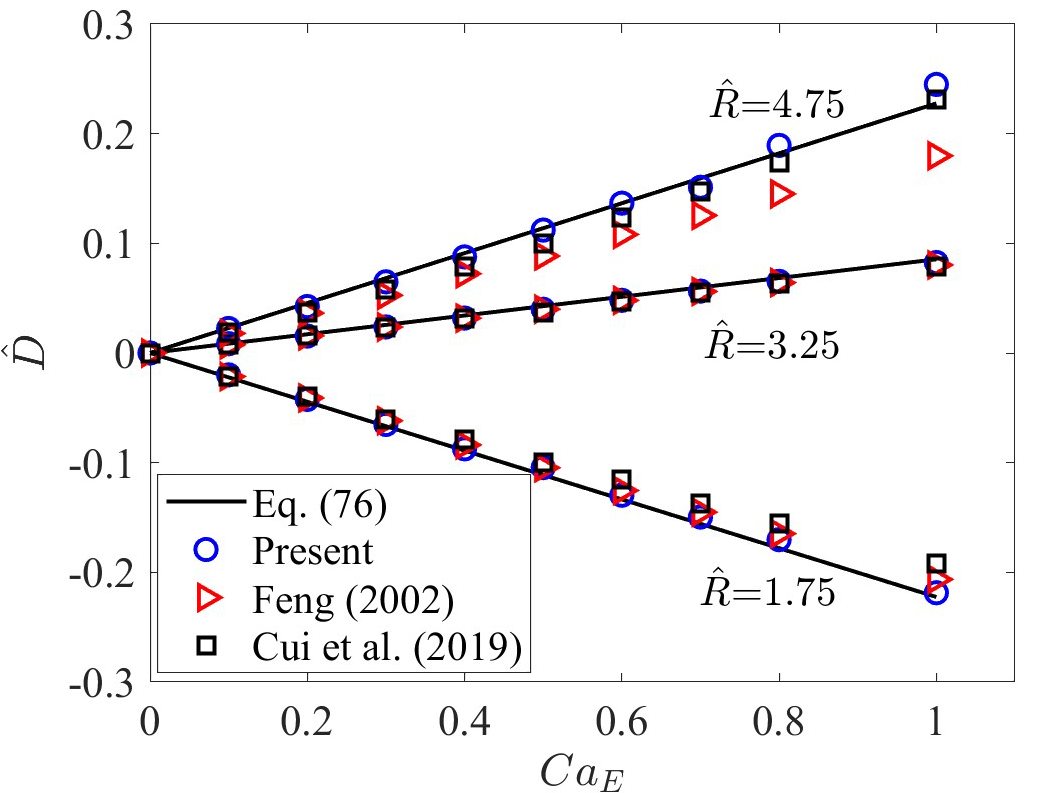}}		
	\subfigure[]{
		\label{wfig8b} 
		\includegraphics[width=0.45\textwidth]{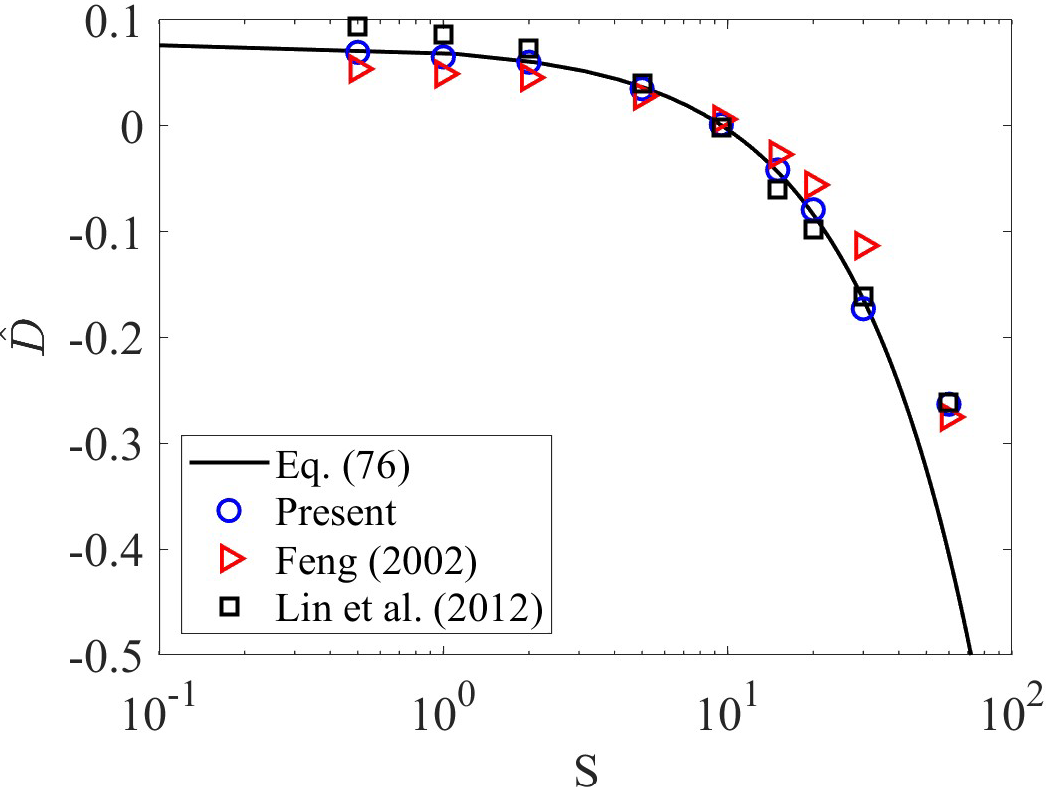}}	
	
	\caption{(a) Effect of electric capillary number $Ca_E$ on droplet deformation $\hat D$ at different conductivity ratios $\hat R$ with $\hat S=3.5$; (b) effect of different permittivity ratios on deformation factor with $Ca_E=0.2$ and $\hat R=5.0$.}
	\label{wfig8}
\end{figure}

To quantify the results, Table. \ref{tab:tb1} presents the deformation factor of the droplet at various parameters, and it can be seen that our numerical results fit well with these existing data  \cite{Luo2020Num} . Moreover, in order to provide a more convincing evaluation, we also conduct additional simulations using various electric capillary numbers, permittivity ratios, and conductivity ratios. We then plot the results in Fig. \ref{wfig8} for comparison. It is evident from this figure that the numerical results from our current model align well with both existing theoretical and numerical benchmark data. These results demonstrate the viability of the proposed approach for droplet deformation in an electric field and indicate that the present model can accurately simulate two-phase EHD flows. 

\subsection{Droplet detachment in reversed electrowetting}

\begin{figure}[H]
	\centering
	\includegraphics[width=0.9\textwidth]{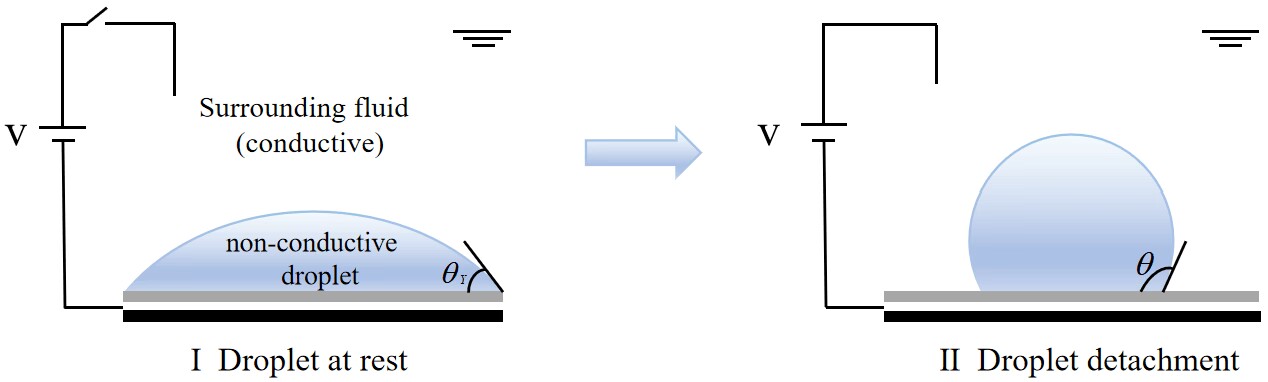}				
	
	\caption{Schematic diagram of reversed electrowetting, in which a nonconducting droplet is immersed in another conducting fluid.}
	\label{wfig9}
\end{figure}

In classical electrowetting experiments, the Lippmann equation is often used to describe the wetting behaviour of a conductive droplet. According to this equation, the contact angle of the droplet decreases as the applied voltage increases. In this setting, when it comes to removing droplets from a solid surface, a critical process in applications such as enhanced oil recovery \cite{Weng2021the}  and surface cleaning \cite{Zhao2008drop} , the only way to do so is by suddenly releasing the applied voltage  \cite{Lee2014ele}. However, due to the strict critical conditions for droplet detachment, this strategy would fail in many conditions, such as the applied voltage being too low or the process not being fast enough to restore sufficient energy  \cite{Vo2019cri} . To this end, recently, some researchers manipulated the non-conductive droplets by using a "reversed" electrowetting effect \cite{Wang2020act} . The setup of this new approach is depicted in Fig. \ref{wfig9}, where a droplet is initially placed on the bottom wall and submerged in water. The droplet's working liquid is a non-conductive fluid, similar to silicone oil. When a potential difference is applied between the bottom wall and the surrounding water, the apparent contact angle of the droplet tends to increase, leading to the contact radius decrease. In this manner, the droplet is expected to separate from the substrate as long as the applied voltage is sufficiently large. Owing to the wettability of the substrate in this setup being increased in applied voltage, the phenomenon is referred to as reversed electrowetting.

We now turn to simulate the detachment of a non-conductivity droplet in reversed electrowetting. Our simulations are performed on a computational domain of $200 \times 100$, and a semicircular droplet with a radius of 25 is initialized in contact with the bottom wall. The density ratio, dynamic viscosity ratio and the electric permittivity ratio are chosen as ${{{{\rho _l}} \mathord{\left/{\vphantom {{{\rho _l}} \rho }} \right.\kern-\nulldelimiterspace} \rho }_v} = 1.069$, ${{{{\mu _l}} \mathord{\left/{\vphantom {{{\mu _l}} \mu }} \right.\kern-\nulldelimiterspace} \mu }_v} = 0.097$, and ${{{{\varepsilon _l}} \mathord{\left/	{\vphantom {{{\varepsilon _l}} \varepsilon }} \right.\kern-\nulldelimiterspace} \varepsilon }_v} = 32.0$, which approach those of realistic water-oil system adopted in the experiment \cite{Weng2021the}. The initial droplet contact angle in the absence of voltage is set to be ${40^ \circ }$. Different electric potentials are imposed between the top (${\varphi _H}$) and bottom walls (${\varphi _L}$), where ${\varphi _H}>{\varphi _L}$. The boundary conditions of the mesoscopic particles, as well as the other unmentioned parameters, are the same as those adopted in Sec. \ref{sect4}. Moreover,  to realize the reversed electrowetting phenomenon, the effective surface tension for liquid-solid should modified as  ${{\hat \gamma }_{sl}} = {\gamma _{sl}} + {\eta  \mathord{\left/{\vphantom {\eta  d}} \right.
\kern-\nulldelimiterspace} d}$.            

\begin{figure}[H]
	\centering
	\subfigure[]{
		\label{wfig10a} 
		\includegraphics[width=0.9\textwidth]{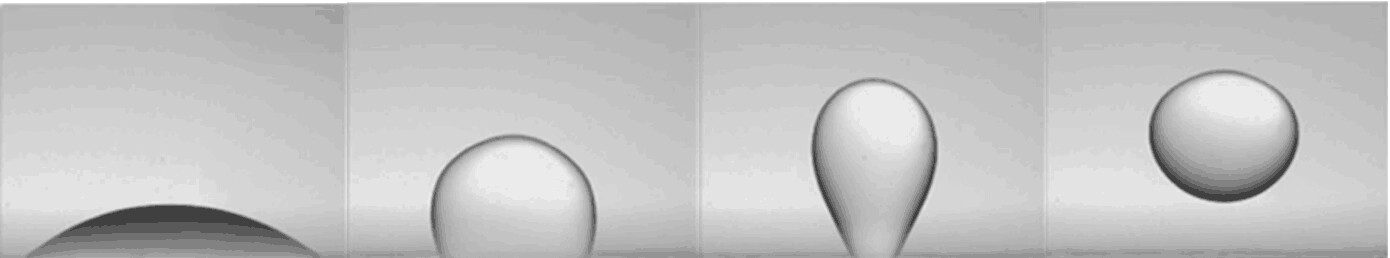}}		
	\subfigure[]{
		\label{wfig10b} 
		\includegraphics[width=0.9\textwidth]{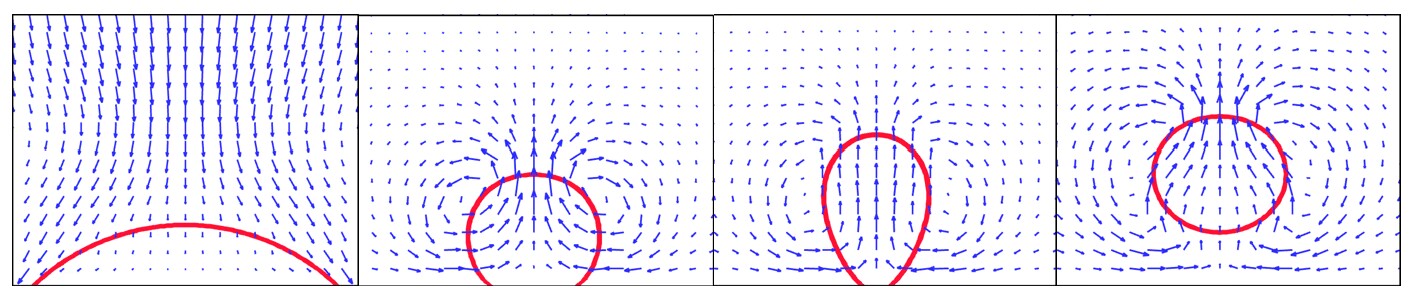}}	
	
	\caption{Comparison of shapes during the process of the reversed electrowetting drives a droplet detachment from the substrate: (a) experimental results \cite{Weng2021the}; (b) the present numerical results (the blue arrows represent the direction of velocity.} 
	\label{wfig10}
\end{figure}

\begin{figure}[H]
	\centering
	\subfigure[]{
		\label{wfig11a} 
		\includegraphics[width=0.48\textwidth]{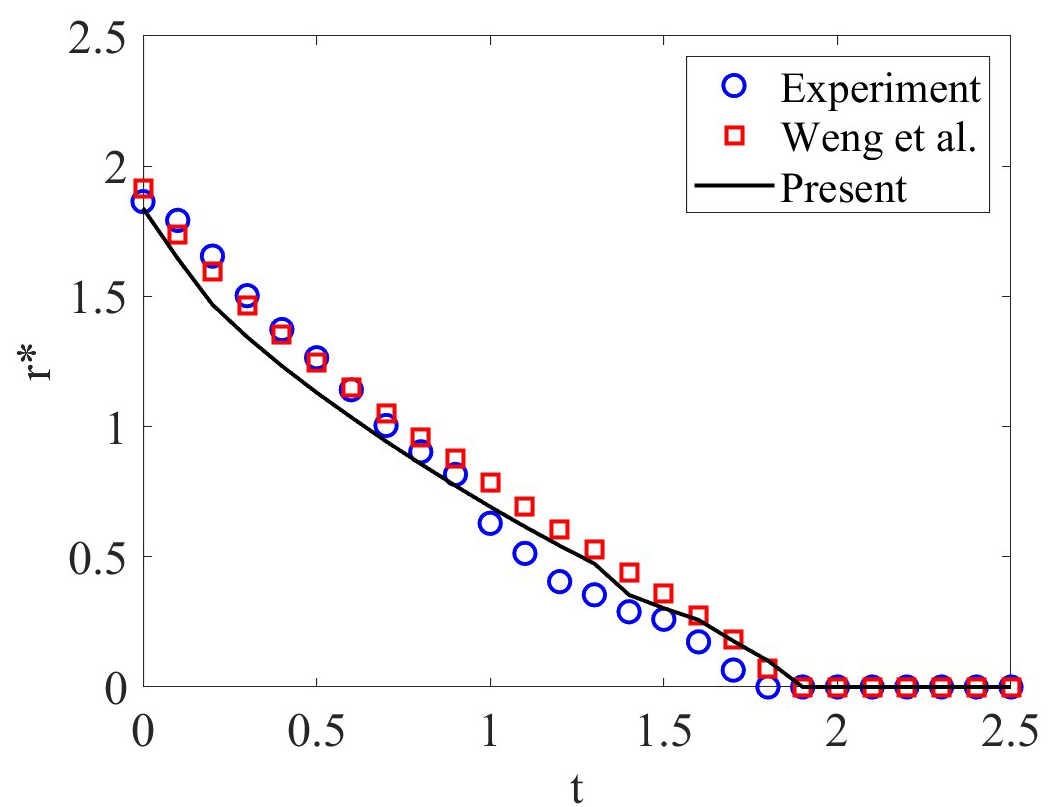}}		
	\subfigure[]{
		\label{wfig11b} 
		\includegraphics[width=0.48\textwidth]{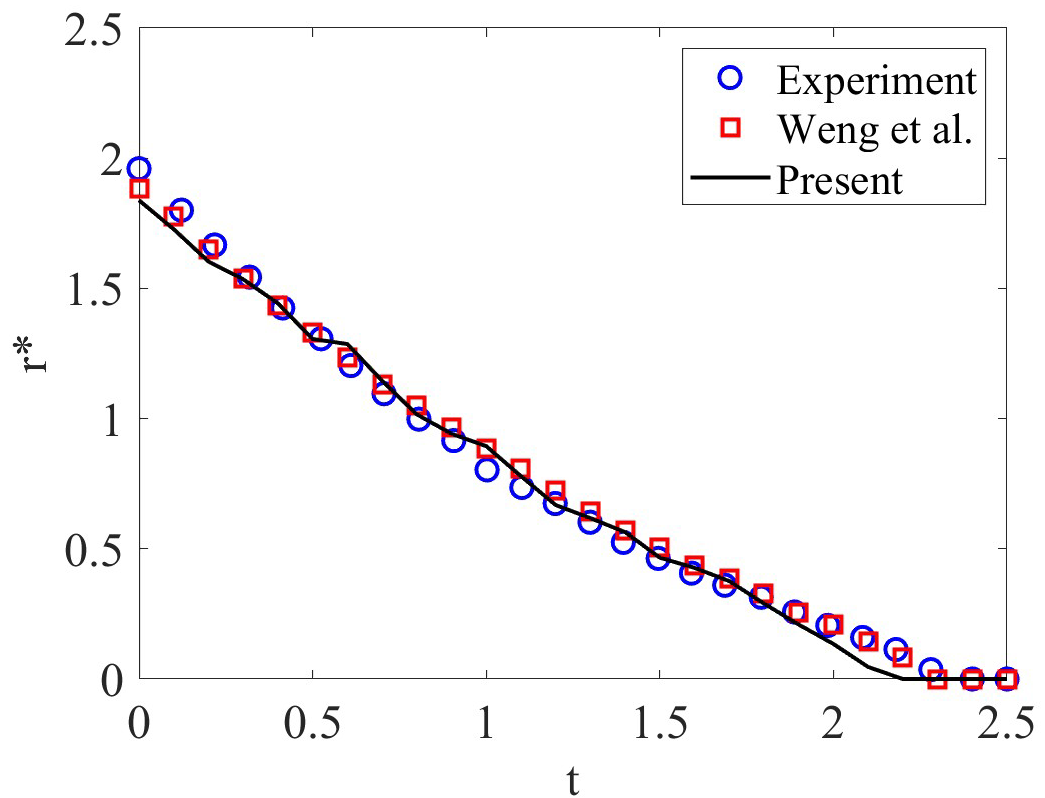}}	
	
	\caption{Evolution of the normalized contact radius on the substrate over time. The present numerical results (black lines) are compared with the experimental results (blue circles) and the available data (red squares): (a) ${{\mathop \mu \nolimits_l } \mathord{\left/{\vphantom {{\mathop \mu \nolimits_l } {\mathop \mu \nolimits_v }}} \right.\kern-\nulldelimiterspace} {\mathop \mu \nolimits_v }} = 0.097$; (b) ${{\mathop \mu \nolimits_l } \mathord{\left/{\vphantom {{\mathop \mu \nolimits_l } {\mathop \mu \nolimits_v }}} \right.\kern-\nulldelimiterspace} {\mathop \mu \nolimits_v }} = 0.022$.} 
	\label{wfig11}
\end{figure}

The comparison of the current simulation with the experiment conducted by Weng et al. \cite{Weng2021the}  is shown in Fig. \ref{wfig10}, which illustrates the variation of the droplet shape and contact radius. As seen from this figure, our simulated results are comparable to those of Weng et al.'s experiment  \cite{Weng2021the} . In particular, we find that in the reversed electrowetting, the droplet tends to contract under the action of the electric field. Owing to the storing surface energy of the droplet being large enough, leading to detachment from the solid surface. Fig. \ref{wfig10b} also depicts the corresponding velocity vectors near the droplet, and it can be seen that there are two counter-rotating vortexes at the front of the droplet interface. For a quantitative analysis, we recorded the time evolution of the contact radius and compared it with experimental studies and numerical data computed by COMSOL. As shown in Fig. \ref{wfig11}, our results are consistent with previous observations, indicating that our model is a suitable candidate for simulating two-phase EHD flows.

\section{Conclusion}
In this paper, we first use Onsager's variational principle to develop a thermodynamically consistent phase-field model for two-phase EHD flows. This model allows a two-phase incompressible fluid to have different electrohydrodynamic properties for each phase, such as densities, viscosities, permittivities, and conductivities. In particular, the deduced model incorporates the influence of the surface charge convection, which is often overlooked in previous studies. This feature enables the model to be applied to problems where bulk charge conduction and convection are relevant, such as the characterization of the cone-to-jet transition region in EHD cone-jet electrosprays.         

After obtaining the thermodynamically consistent phase-field model, the lattice Boltzmann method is utilized to simulate the two-phase EHD flows. In this approach, it adopts four distribution functions for solving the Cahn-Hilliard equation, the electric potential equation, the Nernst-Planck equation, and the hydrodynamic equations. Additionally, we have developed an improved LB method for the electric potential equation to account for two-phase EHD flows with a larger permittivity ratio. The performance of the LB method is evaluated by simulating EHD flows with two superimposed planar fluids and the charge diffusion of a Gaussian bell. The numerical results indicate that the developed LB method has a second-order convergence rate in predicting electric potential and charge density. Moreover, we consider the equilibrium interface profiles of a conductivity droplet and found that the resulting apparent contact agrees well with those derived by the Lippmann equation. To further test the current LB method's capability in practical problems, we also conduct some simulations for two typical two-phase EHD problems, including droplet deformation under an electric field and droplet detachment in reversed electrowetting. Based on the comparison between our numerical results and the analytical solutions, as well as experimental/existing numerical data,  it is found that the present method enables us to derive comparative results for predicting droplet deformation and contact radius. In conclusion, the simulated results demonstrate that the current method is feasible for simulating two-phase EHD flows.

\section*{CRediT authorship contribution statement}
{\bf{Fang Xiong}}: Writing – original draft, Visualization, Validation, Software, Methodology, Investigation, Formal analysis, Data curation. {\bf{Lei Wang}}:  Writing – review \& editing, Writing – original draft, Supervision, Conceptualization. {\bf{Jiangxu Huang}}: Supervision, Software, Methodology, Resources. {\bf{Kang Luo}}: Supervision, Software, Methodology, Funding acquisition.

\section*{Declaration of competing interest}
The authors declare that they have no known competing financial interests or personal relationships that could have appeared to influence the work reported in this paper.

\section*{Data availability}
Data will be made available on request.

\section*{Acknowledgements}
This work is financially supported by the National Natural Science Foundation of China (Grant No. 51906051). 

\appendix

\section{Chapman-Enskog analysis on electric potential equation}
To show that the electric potential equation Eq. (\ref{weq32}) can be recovered with no deviation term using the proposed LB equation Eq. (\ref{weq53}), the Chapman-Enskog analysis is performed in this appendix. Base on the definitions of discrete lattice velocity $\mathbf{c}_i$ and the equilibrium distribution function $ h_i^{eq} $ given by Eq. (\ref{weq37}) and Eq. (\ref{weq54}), we can easily obtain
\begin{equation}
	\sum\limits_i {h_i^{eq} } = 0, \qquad \sum\limits_i {\mathbf{c}_i h_i^{eq} }  = 0, \qquad \sum\limits_i {\mathbf{c}_i \mathbf{c}_i h_i^{eq} }  = c_s^2 \varphi. 
	\label{eqA.3}
\end{equation}
In addition, the distribution function, the derivatives of time and space, as well as the source term can be expanded in consecutive scales of $\xi$:
\begin{equation}
	\begin{gathered}
		h_i  = h_i^{(0)}  + \xi h_i^{(1)}  + \mathop \xi \nolimits^2 h_i^{(2)} + \cdots,  \\
		\frac{\partial }{{\partial t}} = \xi \frac{\partial }{{\partial t_1 }} + \mathop \xi \nolimits^2 \frac{\partial }{{\partial t_2 }} ,\qquad \nabla  = \xi \mathop \nabla \nolimits_1 , \qquad q = \mathop \xi \nolimits^2 \mathop q\nolimits^{(2)} ,
		\label{eqA.4}	 
	\end{gathered}
\end{equation}
where $ \xi $ is the expansion parameter. Then, applying the Taylor expansion to Eq. (\ref{weq53}) at time $t$ and space $\mathbf{x}$, we get
\begin{equation}
	\Delta t D_i h_i  + \frac{{\mathop {\Delta t}\nolimits^2 }}{2} D_i^2 h_i  =  - \frac{1}{{\tau_h }}[h_i  - h_i^{eq} ] - \frac{\hat \varepsilon \left( \phi  \right)}{{\mathop c\nolimits_s^2 }}\frac{{\mathop \omega \nolimits_i \mathbf{c}_i {\nabla_1} \varphi}}{{\tau_h }} {\rm{ + }}\Delta t {\varpi _i } q,
	\label{eqA.5}
\end{equation}
where $ \mathop D\nolimits_i  = \xi \mathop D\nolimits_{1i}  + \mathop \xi \nolimits^2 \mathop \partial \nolimits_{\mathop t\nolimits_2 } $ with $ \mathop D\nolimits_{1i}  = \mathop \partial \nolimits_{\mathop t\nolimits_1 }  + \mathbf{c}_i  \cdot \mathop \nabla \nolimits_1 $. Substituting Eq. (\ref{eqA.4}) into Eq. (\ref{eqA.5}) yields the following equations in the successive order of 
\begin{align}
	& O(\mathop \xi \nolimits^0 ):\mathop h\nolimits_i^{(0)}  = \mathop h\nolimits_i^{eq},  \label{eqA.6} \\
	& O(\mathop \xi \nolimits^1 ):\mathop D\nolimits_{1i} \mathop h\nolimits_i^{(0)}  =  - \frac{1}{{\mathop \tau \nolimits_h \Delta t}}\mathop h\nolimits_i^{(1)}  - \frac{\hat \varepsilon \left( \phi  \right)}{{\mathop c\nolimits_s^2 \Delta t }}\frac{{\mathop \omega \nolimits_i \mathbf{c}_i \nabla_1 \varphi}}{{\tau_h }}, \label{eqA.7} \\
	& O(\mathop \xi \nolimits^2 ):\mathop \partial \nolimits_{\mathop t\nolimits_2 } \mathop h\nolimits_i^{(0)}  + \mathop D\nolimits_{1i} \mathop h\nolimits_i^{(1)}  + \frac{{\Delta t}}{2}\mathop {\mathop D\nolimits_{1i} }\nolimits^2 \mathop h\nolimits_i^{(0)}  =  - \frac{1}{{\mathop \tau \nolimits_h \Delta t}}\mathop h\nolimits_i^{(2)}  + {\varpi _i } \mathop q\nolimits^{(2)}. \label{eqA.8}
\end{align}
Combining Eq. (\ref{eqA.3}), Eq. (\ref{eqA.4}) and Eq. (\ref{eqA.6}), we get
\begin{equation}
	\sum\limits_i {\mathop h\nolimits_i^{(k)} }  = 0 \quad (k > 1).
	\label{eqA.9}
\end{equation}
Moreover, applying Eq. (\ref{eqA.7}) to Eq. (\ref{eqA.8}), one can obtain
\begin{equation}
	\mathop \partial \nolimits_{\mathop t\nolimits_2 } \mathop h\nolimits_i^{(0)}  + \mathop D\nolimits_{1i} (1 - \frac{1}{{2\mathop \tau \nolimits_h }})\mathop h\nolimits_i^{(1)}  - \mathop D\nolimits_{1i} \frac{\hat \varepsilon \left( \phi  \right)}{{\mathop {2c}\nolimits_s^2 }}\frac{{\mathop \omega \nolimits_i \mathbf{c}_i \mathop \nabla \nolimits_1 \varphi }}{{\mathop \tau \nolimits_h }} =  - \frac{1}{{\mathop \tau \nolimits_h \Delta t}}\mathop h\nolimits_i^{(2)}  + {\varpi _i } \mathop q\nolimits^{(2)} .
	\label{eqA.10}
\end{equation}
After a summation of Eq. (\ref{eqA.10}) over $ i $, we can get the scale equation as
\begin{equation}
	(1 - \frac{1}{{2\mathop \tau \nolimits_h }})\mathop \nabla \nolimits_1 \sum\limits_i {\mathbf{c}_i } \mathop h\nolimits_i^{(1)}  - \mathop \nabla \nolimits_1 \frac{\hat \varepsilon \left( \phi  \right)}{{2\mathop \tau \nolimits_h }}\mathop \nabla \nolimits_1 \varphi  = \mathop q\nolimits^{(2)} .
	\label{eqA.11}
\end{equation}
On the basis of Eq. (\ref{eqA.7}), we have
\begin{equation}
	\begin{split}
		\sum\limits_i {\mathbf{c}_i } \mathop h\nolimits_i^{(1)}  &= \mathop \tau \nolimits_h \Delta t[ - \frac{\hat \varepsilon \left( \phi  \right)}{{\mathop c\nolimits_s^2 \Delta t\mathop \tau \nolimits_h }}\sum\limits_i {\mathop \omega \nolimits_i \mathbf{c}_i \mathbf{c}_i {\mathop \nabla \nolimits_1} \varphi }  - \mathop \nabla \nolimits_1 \sum\limits_i {\mathbf{c}_i \mathbf{c}_i \mathop h\nolimits_i^{(0)} }  - \mathop \partial \nolimits_{\mathop t\nolimits_1 } \sum\limits_i {\mathbf{c}_i \mathop h\nolimits_i^{(0)} } ] \\
		&=  - {\hat \varepsilon \left( \phi  \right)}{\mathop \nabla \nolimits_1}  \varphi  - \mathop \tau \nolimits_h \Delta t\mathop c\nolimits_s^2 \mathop \nabla \nolimits_1 \varphi.
	\end{split}
	\label{eqA.12}
\end{equation}
With the aid of the above equation, Eq. (\ref{eqA.11}) can be rewritten as 
\begin{equation}
	\mathop \nabla \nolimits_1 [\mathop c\nolimits_s^2 (\mathop \tau \nolimits_h  - \frac{1}{2})\Delta t\mathop \nabla \nolimits_1 \varphi ] =  - \mathop q\nolimits^{(2)}  - \mathop \nabla \nolimits_1 \mathop {(\hat \varepsilon \left( \phi  \right) \mathop \nabla \nolimits_1 \varphi } ).
	\label{eqA.13}
\end{equation}
Multiplying $\mathop \xi \nolimits^2 $ on both sides of Eq. (\ref{eqA.13}), we can derive the electric potential equation Eq. (\ref{weq52}) with $\varepsilon_v$ being 
\begin{equation}
	\mathop \varepsilon \nolimits_v \mathop { = c}\nolimits_s^2 (\mathop \tau \nolimits_h  - \frac{1}{2})\Delta t.
	\label{eqA.15}
\end{equation}
Particularly, according to Eq. (\ref{eqA.7}), we obtain a local scheme for computing $	\nabla \varphi $, which is expressed as  
\begin{equation}
	\nabla \varphi  =  - \frac{{\sum\limits_i {\mathbf{c}_i} \mathop h\nolimits_i^{(1)} }}{{\hat \varepsilon \left( \phi  \right) + \mathop c\nolimits_s^2 \mathop \tau \nolimits_h \Delta t}} =  - \frac{{\sum\limits_i {\mathbf{c}_i } \mathop h\nolimits_i }}{{\hat \varepsilon \left( \phi  \right) + \mathop c\nolimits_s^2 \mathop \tau \nolimits_h \Delta t}}.
	\label{eqA.16}
\end{equation}

\section{Chapman-Enskog analysis on Nernst-Planck equation}
Similar to the process derivation, the equilibrium distribution function for Nernst-Planck equation satisfy 
\begin{equation}
	\sum\limits_i {l_i^{eq} } = q, \qquad \sum\limits_i {\mathbf{c}_i l_i^{eq} }  = q\mathbf{u}, \qquad \sum\limits_i {\mathbf{c}_i \mathbf{c}_i l_i^{eq} }  = c_s^2 q. 
	\label{eqB.1}
\end{equation}
Applying the Taylor series expansion to Eq. (\ref{weq58}), we have
\begin{equation}
	\Delta t\mathop D\nolimits_i \mathop l\nolimits_i  + \frac{{\mathop {\Delta t}\nolimits^2 }}{2}\mathop {\mathop D\nolimits_i }\nolimits^2 \mathop l\nolimits_i =  - \frac{1}{{\mathop \tau \nolimits_l }}(\mathop l\nolimits_i  - \mathop l\nolimits_i^{eq} ) + \Delta t\mathop S\nolimits_i  + \Delta t\mathop T\nolimits_i . 
	\label{eqB.5}
\end{equation}
By introducing the following expansions
\begin{align}
	& \mathop l\nolimits_i  = \mathop l\nolimits_i^{(0)}  + \xi \mathop l\nolimits_i^{(1)}  + \mathop \xi \nolimits^2 \mathop l\nolimits_i^{(2)}  +  \cdots,  \label{eqB.2} \\
	& \mathop \partial \nolimits_t  = \varepsilon \mathop \partial \nolimits_{\mathop t\nolimits_1 }  + \mathop \xi \nolimits^2 \mathop \partial \nolimits_{\mathop t\nolimits_2 } ,\qquad \nabla  = \xi \mathop \nabla \nolimits_1 , \label{eqB.3} \\
	& S = \xi \mathop S\nolimits^{(1)}, \qquad T = \xi \mathop T\nolimits^{(1)}  + \mathop \xi \nolimits^2 \mathop T\nolimits^{(2)}, \label{eqB.4}
\end{align}
and substituting Eqs. (\ref{eqB.2})-(\ref{eqB.4}) into  Eq. (\ref{eqB.5}) yields 
\begin{align}
	& O(\mathop \xi \nolimits^0 ):\mathop l\nolimits_i  = \mathop l\nolimits_i^{eq},  \label{eqB.6} \\
	& O(\mathop \xi \nolimits^1 ):\mathop D\nolimits_{1i} \mathop l\nolimits_i^{(0)}  =  - \frac{1}{{\Delta t\mathop \tau \nolimits_l }}\mathop l\nolimits_i^{(1)}  + \mathop S\nolimits_i^{(1)}  + \mathop T\nolimits_i^{(1)} , \label{eqB.7}\\
	& O(\mathop \xi \nolimits^2 ):\mathop \partial \nolimits_{\mathop t\nolimits_2 } \mathop l\nolimits_i^{(0)}  + \mathop D\nolimits_{1i} \mathop l\nolimits_i^{(1)}  + \frac{{\Delta t}}{2}\mathop {\mathop D\nolimits_{1i} }\nolimits^2 \mathop l\nolimits_i^{(0)}  =  - \frac{1}{{\Delta t\mathop \tau \nolimits_l }}\mathop l\nolimits_i^{(2)}  + \mathop T\nolimits_i^{(2)}. \label{eqB.8}
\end{align}
With Eq. (\ref{weq62}), Eq. (\ref{eqB.1}) and Eq. (\ref{eqB.2}), one can obtain
\begin{align}
	& \sum\limits_i {\mathop l\nolimits_i^{(1)}  =  - \frac{{\Delta t}}{2}} R,  \label{eqB.9} \\
	& \sum\limits_i {\mathop l\nolimits_i^{(k)}  = 0, \quad k > 1} . \label{eqB.10}
\end{align}
where the forcing term $R$ is given by Eq. (\ref{weq57}) .
Combining  Eq. (\ref{eqB.7}) and Eq. (\ref{eqB.8}), we get 
\begin{equation}
	\mathop \partial \nolimits_{\mathop t\nolimits_2 } \mathop l\nolimits_i^{(0)}  + (1 - \frac{1}{{2\mathop \tau \nolimits_l }})\mathop D\nolimits_{1i} \mathop l\nolimits_i^{(1)}  + \frac{{\Delta t}}{2}\mathop D\nolimits_{1i} (\mathop S\nolimits_i^{(1)}  + \mathop T\nolimits_i^{(1)} ) =  - \frac{1}{{\Delta t\mathop \tau \nolimits_l }}\mathop l\nolimits_i^{(2)}  + \mathop T\nolimits_i^{(2)}.  
	\label{eqB.11}
\end{equation}
Then, summing Eq. (\ref{eqB.7})  and Eq. (\ref{eqB.11}) over $i$, we have 
\begin{align}
	& \mathop \partial \nolimits_{\mathop t\nolimits_1 } q + \mathop \nabla \nolimits_1  \cdot (q\mathbf{u}) = R,  \label{eqB.12} \\
	& \mathop \partial \nolimits_{\mathop t\nolimits_2 } q + \mathop \nabla \nolimits_1 [(\frac{1}{2} - \mathop \tau \nolimits_l )\Delta t\mathop c\nolimits_S^2 \mathop \nabla \nolimits_1 q] = 0. \label{eqB.13}
\end{align}
Eventually, taking Eq. (\ref{eqB.12}) $\times  \xi$ + Eq. (\ref{eqB.13}) $\times \mathop \xi \nolimits^2$ supplemented with the imcompressibility condition $\nabla  \cdot \mathbf{u} = 0$, the Nernst-Planck equation can be derived with
\begin{equation}
	\alpha = (\mathop \tau \nolimits_l  - \frac{1}{2})\mathop c\nolimits_S^2 \Delta t.
	\label{eqB.15}
\end{equation}


\end{document}